\documentclass[%
 aip,
 amsmath,amssymb,
 reprint,
]{revtex4-1}

\usepackage{graphicx}
\usepackage{dcolumn}
\usepackage{bm}

\usepackage[utf8]{inputenc}
\usepackage[T1]{fontenc}
\usepackage{mathptmx}
\usepackage{color}

\begin{document}

\preprint{AIP/123-QED}


\title[ ]{Theory and optimisation of radiative recombination in broken-gap InAs/GaSb superlattices}

\author{C\'{o}nal Murphy}
 \email{conalmurphy@umail.ucc.ie}
 \affiliation{Tyndall National Institute, University College Cork, Lee Maltings, Dyke Parade, Cork T12 R5CP, Ireland}
 \affiliation{School of Physics, University College Cork, Cork T12 YN60, Ireland}

\author{Eoin P.~O'Reilly}
 \affiliation{Tyndall National Institute, University College Cork, Lee Maltings, Dyke Parade, Cork T12 R5CP, Ireland}
 \affiliation{School of Physics, University College Cork, Cork T12 YN60, Ireland}

\author{Christopher A.~Broderick}
 \email{c.broderick@umail.ucc.ie}
 \affiliation{Tyndall National Institute, University College Cork, Lee Maltings, Dyke Parade, Cork T12 R5CP, Ireland}
 \affiliation{School of Physics, University College Cork, Cork T12 YN60, Ireland}

\date{\today}


\begin{abstract}

We present a theoretical analysis of mid-infrared radiative recombination in InAs/GaSb superlattices (SLs). We employ a semi-analytical plane wave expansion method in conjunction with an 8-band $\mathbf{k} \cdot \mathbf{p}$ Hamiltonian to compute the SL electronic structure, paying careful attention to the identification and mitigation of spurious solutions. The calculated SL eigenstates are used directly to compute spontaneous emission spectra and the radiative recombination coefficient $B$. We elucidate the origin of the relatively large $B$ coefficients in InAs/GaSb SLs which, despite the presence of spatially indirect (type-II-like) carrier confinement, are close to that of bulk InAs and compare favourably to those calculated for mid-infrared type-I pseudomorphic and metamorphic quantum well structures having comparable emission wavelengths. Our analysis explicitly quantifies the roles played by carrier localisation (specifically, partial delocalisation of bound electron states) and miniband formation (specifically, miniband occupation and optical selection rules) in determining the magnitude of $B$ and its temperature dependence. We perform a high-throughput optimisation of the room temperature $B$ coefficient in InAs/GaSb SLs across the 3.5 -- 7 $\mu$m wavelength range, quantifying the dependence of $B$ on the relative thickness of the electron-confining InAs and hole-confining GaSb layers. This analysis provides guidance for the growth of optimised SLs for mid-infrared light emitters. Our results, combined with the expected low non-radiative Auger recombination rates in structures having spatially indirect electron and hole confinement, corroborate recently observed high output power in prototype InAs/GaSb SL inter-band cascade light-emitting diodes.

\end{abstract}

\maketitle


\section{Introduction}


The narrow-gap semiconductors InAs and GaSb, members of the so-called ``6.1 \AA~family'' of III-V compounds, \cite{Kroemer_PE_2004} combine low lattice mismatch with a large valence band (VB) offset that produces an unusual type-III (``broken-gap'') band alignment. This low lattice mismatch allows for growth of high-quality superlattices (SLs) consisting of alternating InAs and GaSb layers, and has established InAs/GaSb SLs as the archetypal broken-gap SL system. Research effort on InAs/GaSb SLs and quantum wells (QWs) has been sustained since their initial fabrication in the late-1970s \cite{Sakaki_SSC_1978} by the vast possibilities offered by their highly tunable electronic structure, which continues to attract interest from both fundamental and applied perspectives. From a fundamental perspective InAs/GaSb SLs have long served as a testbed system for electronic structure methodologies, including empirical atomistic \cite{Ihm_PRB_1979,Grosso_PRB_1989,Wei_PRB_2004,Mir_JAP_2013,Kato_SM_2018,Sawamura_OME_2018} and continuum ($\mathbf{k} \cdot \mathbf{p}$-based) \cite{Bastard_PRB_1981,Bastard_PRB_1982,Altarelli_PRB_1983,Johnson_PRB_1990} models and, more recently, first principles calculations. \cite{Wang_JAP_2014,Otsuka_JJAP_2017,Garwood_IPT_2017,Taghipour_JPCM_2018,Yang_PRM_2021} In addition to sustained interest from the perspective of mid-infrared photonics, InAs/GaSb heterostructures have over the past 15 years also attracted an ongoing surge of interest related to the ability to engineer topologically non-trivial electronic states, \cite{Liu_PRL_2008,Knez_PRL_2011,Qu_PRL_2015} including the pursuit of Majorana fermions as a platform for fault-tolerant quantum computation. \cite{Pribiag_NN_2015}


From a practical perspective, there exists a strong imperative to develop efficient and cost-effective mid-infrared light sources for a broad range of sensing applications relevant to the environmental, industrial, medical, agricultural and defence sectors. \cite{Krier_Springer_mid-IR,Ilev_Springer_2006,Tournie_Elsevier_2012,Wang_OSA_2016} While quantum cascade laser (QCL) technology is well-established, \cite{Tournie_Elsevier_2012,Vitiello_OSA_2015,Razeghi_OSA_2015} QCL performance degrades for wavelengths $\lesssim 4$ $\mu$m and their deployment presents the dual challenges of complex fabrication and high cost. This represents a significant opportunity for the development of novel light-emitting diodes (LEDs) and diode lasers based on III-V semiconductor heterostructures. Such devices have the potential to deliver simplified fabrication, lower cost and reduced power consumption compared to QCL technologies, and to significantly improve performance compared to existing incandescent light sources. \cite{Krier_chapter_2020} Mid-infrared emitters based on QWs having type-I band offsets experience performance degradation with increasing wavelength, due to a combination of thermal carrier leakage and high non-radiative Auger recombination rates. \cite{Sifferman_IEEE_2015,Eales_IEEE_2017} Heterostructures having type-II (staggered-gap) band offsets have attracted significant attention for the development of mid-infrared emitters. \cite{Meyer_APL_1995,Ricker_JAP_2017,Delli_APL_2020,Meyer_IEEEJQE_2021,Tournie_LSA_2022} Compared to QWs or SLs having type-I (spatially direct) band offsets, they offer a combination of enhanced electrical confinement and intrinsically low Auger recombination rates.\cite{Zegrya_APL_1995} This can enhance internal quantum efficiency by respectively mitigating carrier leakage and reducing non-radiative losses. \cite{Muhowski_JAP_2019,Muhowski_APL_2020} InAs/GaSb SLs, already the basis of demonstrated mid-infrared photodetectors, \cite{Mohseni_APL_2001,Wei_APL_2005,Zhang_IEEE_2011} have continued to attract interest as a candidate for the development of mid-infrared LEDs. Progress on light-emitting InAs/GaSb SLs has included fabrication of prototype inter-band cascade LEDs (IC-LEDs) by several groups. \cite{Koerperick_IEEE_2009,Koerperick_IEEE_2011,Provence_JAP_2015} These prototypes have demonstrated low turn-on voltage, low series resistance, and room temperature radiance approaching 1 W cm$^{-2}$ sr$^{-1}$. \cite{Zhou_APL_2019}


In this paper we undertake a detailed analysis of the electronic structure of, and radiative recombination in, InAs/GaSb SLs. Recombination processes in these structures have been the subject of previous theoretical analysis. \cite{Grein_JAP_2002} However, to our knowledge, there has not been (i) detailed discussion of the explicit role played by various aspects of the underlying electronic structure in determining the radiative recombination rate and its temperature dependence, or (ii) systematic in silico optimisation of the radiative recombination rate as a function of emission wavelength. It is to these two issues that we dedicate our attention. We recapitulate the calculation of SL electronic structure using a multi-band $\mathbf{k} \cdot \mathbf{p}$ Hamiltonian, employing here a reciprocal space plane wave expansion method (PWEM). This numerically efficient approach allows the miniband dispersion to be computed explicitly using a calculational supercell consisting of a single SL period. We highlight that analysis of the bulk \textit{complex} band structure is essential to identify and inform mitigation of spurious solutions in $\mathbf{k} \cdot \mathbf{p}$-based heterostructure calculations. By revisiting the fundamentals of the InAs/GaSb SL electronic structure we identify and quantify the role played by miniband formation in determining the radiative recombination rate, which we quantify via the radiative recombination coefficient $B$. By comparing the results of our full SL calculations to test calculations for an equivalent QW - i.e.~by neglecting miniband dispersion - we highlight the importance of the mismatch in electron and hole miniband dispersion in influencing the magnitude of $B$ and its temperature dependence. We demonstrate that miniband formation acts to reduce the magnitude of $B$, while simultaneously reducing the rate at which $B$ decreases with increasing temperature, and identify that this behaviour has its origin in the mismatch in dispersion between the highly (minimally) dispersive electron (hole) minibands. This analysis quantifies the potential enhancement in radiative recombination rate that could be achieved by engineering SL band structure to better match electron and hole miniband dispersion. This is in a manner akin to that achieved by exploiting the impact of strain on the VB dispersion in QW lasers, \cite{Reilly_IEEEJQE_1994,Sweeney_JAP_2019} providing guidance to inform structural optimisation that enhances the internal quantum efficiency of SL-based emitters. Finally, we perform high-throughput calculations in which the relative thickness of the InAs and GaSb layers that constitute each period of the SL are varied, identifying combinations of layer thickness that maximise $B$ at room temperature for emission wavelengths in the 3.5 -- 7 $\mu$m range.


The remainder of this paper is organised as follows. Our theoretical model is described in Sec.~\ref{sec:theoretical_model}, beginning in Secs.~\ref{sec:theory_electronic} and~\ref{sec:theory_optical} with the respective details of our SL electronic structure and radiative recombination calculations. Technical considerations related to the mitigation of spurious solutions are described in Sec.~\ref{sec:theory_technical_considerations}. Our results are presented in Sec.~\ref{sec:results}, beginning in Sec.~\ref{sec:results_benchmark} with benchmarking of our electronic structure calculations against experimental data for SLs having equal InAs and GaSb layer thickness. In Sec.~\ref{sec:results_equal_thickness} we elucidate the electronic and optical properties of equal layer thickness SLs, and quantify the impact of miniband formation on radiative recombination. In Sec.~\ref{sec:results_unequal_thickness} we consider InAs/GaSb SLs having unequal InAs and GaSb layer thickness, and undertake systematic optimisation of the radiative recombination coefficient $B$ across the 3.5 -- 7 $\mu$m wavelength range. In Sec.~\ref{sec:conclusions} we summarise and conclude.


\section{Theoretical model}
\label{sec:theoretical_model}


\subsection{Electronic properties: $\mathbf{k} \cdot \mathbf{p}$ Hamiltonian, plane wave expansion method and SL miniband structure}
\label{sec:theory_electronic}


Our calculation of SL electronic properties is based on an 8-band $\mathbf{k} \cdot \mathbf{p}$ Hamiltonian, including spin-orbit coupling, \cite{Cohen_PRB_1990,Bahder_PRB_1990} constructed using a basis of spin-degenerate $\Gamma$-point Bloch functions $\vert u_{b} \rangle$ ($b = 1, 2, \dots, 8$) associated with the lowest energy conduction band (CB), as well as the heavy-hole (HH), light-hole (LH) and spin-split-off (SO) VBs. We consider a [001]-oriented heterostructure in which the growth direction aligns with the $z$ axis of a Cartesian coordinate system, and compute the electronic properties in the envelope function approximation (EFA).

In the EFA, the PWEM consists of expanding the envelope functions $F_{nb\mathbf{k}} (z)$ as a truncated Fourier series, such that a general SL eigenstate $\vert \psi_{n\mathbf{k}} (z) \rangle$ having energy $E_{n\mathbf{k}}$ at wave vector $\mathbf{k} = ( \mathbf{k}_{\parallel}, q )$ is given by \cite{Healy_IEEE_2006,Murphy_NUSOD_2021}

\begin{equation}
    \left| \psi_{n \mathbf{k}} ( z ) \right. \rangle = \sum_{b = 1}^{ N_{b} } \left( \frac{1}{ \sqrt{L} } \, \sum_{m = -M}^{ +M } a_{nbm \mathbf{k} } \, e^{ i \left( G_{m} + q \right) z } \right) \left| u_{b} \right. \rangle \, ,
    \label{eq:plane_wave_expansion}
\end{equation}

\noindent
where $N_{b} = 8$ is the number of Bloch bands included in the underlying bulk $\mathbf{k} \cdot \mathbf{p}$ Hamiltonian, and $\mathbf{k}_{\parallel}$ and $q$ are respectively the components of the wave vector in the plane perpendicular to the growth direction and along the growth direction. The terms in parentheses describe the expansion of the envelope function $F_{nb\mathbf{k}} (z)$ associated with each bulk band $b$ over a supercell $z \in [ - \frac{L}{2}, \frac{L}{2} ]$ of length $L$, in a periodic basis of plane waves having discrete wave vectors $G_{m} = \frac{ 2m \pi }{ L }$ for integers $m \in [ -M, M ]$.

The reciprocal space solution of the associated multi-band EFA Schr\"{o}dinger equation $\widehat{H}_{\mathbf{k}} \vert \psi_{n\mathbf{k}} (z) \rangle = E_{n\mathbf{k}} \vert \psi_{n\mathbf{k}} (z) \rangle$ proceeds via substitution of Eq.~\eqref{eq:plane_wave_expansion} to obtain the reciprocal space matrix representation of the element $\widehat{H}_{b'b,\mathbf{k}}$ of the heterostructure Hamiltonian $\widehat{H}_{\mathbf{k}}$ \cite{Erhardt_Springer_2014,Broderick_thesis_2015}

\begin{equation}
     \left( \widetilde{H}_{b'b,\mathbf{k}} \right)_{m'm} = \frac{1}{L} \int_{ - \frac{L}{2} }^{ + \frac{L}{2} } e^{ - i \left( G_{m'} + q \right) z } \, \widehat{H}_{b'b,\mathbf{k}} (z) \, e^{ i \left( G_{m} + q \right) z } \, \text{d} z \, ,
     \label{eq:plane_wave_hamiltonian}
\end{equation}

\noindent
which is a complex-valued $( 2M + 1 ) \times ( 2M + 1 )$ matrix.

In Eq.~\eqref{eq:plane_wave_hamiltonian}, the operators $\widehat{H}_{b'b,\mathbf{k}} (z)$ must be derived from the bulk $\mathbf{k} \cdot \mathbf{p}$ matrix elements $H_{b'b,\mathbf{k}}$ via (i) wave vector quantisation $k_{z} \to -i \frac{ \text{d} }{ \text{d} z }$, and (ii) application of appropriate boundary conditions to account for the $z$-dependent material parameters. We employ boundary conditions based on symmetrisation of $H_{b'b,\mathbf{k}}$ with respect to $k_{z}$, followed by quantisation of $k_{z}$, which imposes continuity of probability current density across layer interfaces within the supercell. Previous comparative analysis between these widely-employed boundary conditions and the ``exact'' Burt-Foreman EFA boundary conditions, demonstrated that asymmetric contributions in the latter play a minimal role in determining the character of eigenstates computed using an 8-band $\mathbf{k} \cdot \mathbf{p}$ Hamiltonian. \cite{Meney_PRB_1994} We note that $\widehat{H}_{b'b,\mathbf{k}} (z)$ are in general differential operators, so that their matrix representation via Eq.~\eqref{eq:plane_wave_hamiltonian} is $q$-dependent. Symmetrising, quantising and evaluating Eq.~\eqref{eq:plane_wave_hamiltonian} for all matrix elements of the underlying bulk $\mathbf{k} \cdot \mathbf{p}$ Hamiltonian yields the reciprocal space representation $\widetilde{H}_{\mathbf{k}}$ of $\widehat{H}_{\mathbf{k}}$, which is a $N_{b} ( 2M + 1 ) \times N_{b} ( 2M + 1 )$ Hermitian matrix. The eigenvalues of $\widetilde{H}_{\mathbf{k}}$ are the band energies $E_{n\mathbf{k}}$, while its eigenvectors are the Fourier coefficients $a_{nbm \mathbf{k} }$ of the envelope functions $F_{nb \mathbf{k} } (z)$.

The discrete wave vectors $G_{m}$ encapsulate $k_{z}$ quantisation along the growth ($z$) direction. The choice of the plane wave basis states introduces periodicity along the growth direction with period $L$, the miniband dispersion associated with which is described by the wave vector $ -\frac{\pi}{L} \leq q \leq +\frac{\pi}{L} $. Typically, plane wave $\mathbf{k} \cdot \mathbf{p}$ calculations for single- or multi-QW heterostructures are undertaken by neglecting any $q$ dependence -- i.e.~by setting $q = 0$ in Eqs.~\eqref{eq:plane_wave_expansion} and~\eqref{eq:plane_wave_hamiltonian}. This choice is made since (i) thick surrounding barrier layers in QW structures make $L$ large, and hence $q$ small, and (ii) electronic coupling between QW eigenstates localised in neighbouring supercell periods can generally be neglected. Defining a calculational supercell consisting of a single SL period and retaining the explicit $q$ dependence in Eqs.~\eqref{eq:plane_wave_expansion} and~\eqref{eq:plane_wave_hamiltonian} allows to compute the full $\mathbf{k} = ( \mathbf{k}_{\parallel}, q )$ dependent band dispersion and eigenstates, efficiently yielding the electronic structure of an idealised infinite SL. The integrals appearing in Eq.~\eqref{eq:plane_wave_hamiltonian} can be evaluated analytically, so that the accuracy of a given numerical PWEM calculation is therefore limited only by the number $2M + 1$ of plane waves included in the basis set.


While an appropriately parameterised 8-band $\mathbf{k} \cdot \mathbf{p}$ Hamiltonian provides a robust approach to compute SL electronic properties, its artificially high symmetry neglects the impact of microscopic interface asymmetry (MIA) associated with InAs/GaSb layer interfaces. \cite{Erhardt_Springer_2014} To overcome this limitation, previous 8-band $\mathbf{k} \cdot \mathbf{p}$ calculations for InAs/GaSb SLs have treated MIA via inclusion of a localised potential at each interface, the magnitude of which was treated as an empirical parameter and adjusted to fit to the measured variation with layer thickness of the SL band gap. \cite{Li_PRB_2010,Dong_TSF_2015} The primary impact of this symmetry-breaking potential is to induce hybridisation of HH- and LH-like eigenstates, \cite{Rossler_SSC_2002} which can induce strong in-plane anisotropy in the inter-band optical matrix elements. Here, we are concerned with computing the spontaneous emission (SE) rate, and quantities derived therefrom, which are polarisation-averaged. We therefore neglect MIA in the present analysis, and further employ the axial approximation such that the computed in-plane band dispersion and eigenstates depend only on the magnitude $k_{\parallel} = \vert \mathbf{k}_{\parallel} \vert$ of the in-plane wave vector. \cite{Altarelli_PRB_1985}


To compute energy-dependent properties in a QW, e.g., density of states (DOS) and SE, it is necessary to integrate over the in-plane degrees of freedom -- i.e. the in-plane wave vector $\mathbf{k}_{\parallel}$. For a SL this is modified to incorporate $q$-dependent miniband dispersion and eigenstates as

\begin{equation}
    \int \frac{\mathrm{d} \mathbf{k}_{\|}}{(2 \pi)^{2}} \longrightarrow\left(\frac{2 \pi}{L}\right)^{-1} \int_{-\frac{\pi}{L}}^{+\frac{\pi}{L}} \mathrm{~d} q \int \frac{\mathrm{d} \mathbf{k}_{\|}}{(2 \pi)^{2}} \, ,
    \label{eq:integration_conversion}
\end{equation}

\noindent
where $\frac{ 2 \pi }{L}$ is the SL Brillouin zone (BZ) extent along $z$. In the axial approximation we further replace $\text{d} \mathbf{k}_{\parallel}$ by $2 \pi k_{\parallel} \text{d} k_{\parallel}$, reducing the integral of Eq.~\eqref{eq:integration_conversion} from three to two dimensions.


\subsection{Optical properties: optical matrix elements, spontaneous emission spectrum and radiative recombination coefficient}
\label{sec:theory_optical}


To quantify SL radiative performance, we firstly compute the SE spectrum $r_{\text{sp}} ( \hbar \omega )$, where $\hbar \omega$ is the photon energy. This requires evaluation of the inter-band optical (momentum) matrix elements $\mathbf{p}_{n_{c}n_{v}\mathbf{k}}$ between an electron eigenstate $\left| \psi_{n_{c} \mathbf{k}} ( z ) \right. \rangle$ in conduction subband $n_{c}$ and a hole eigenstate $\left| \psi_{n_{v} \mathbf{k}} ( z ) \right\rangle$ in valence subband $n_{v}$, at wave vector $\mathbf{k}$ (where strict $\mathbf{k}$-selection, $\Delta \mathbf{k} = 0$ for optical transitions, imposes that the electron and hole states have equal wave vector). We compute $\mathbf{p}_{n_{c}n_{v}\mathbf{k}}$ via the Hellmann-Feynman theorem, \cite{Szmulowicz_PRB_1995} using the SL Hamiltonian and eigenstates explicitly at each wave vector $\mathbf{k}$

\begin{equation} 
    p_{ n_{c} n_{v} \mathbf{k} }^{ ( e ) } \equiv \widehat{e} \cdot \mathbf{p}_{n_{c}n_{v}\mathbf{k}} = \frac{ m_{0} }{ \hbar } \langle \psi_{ n_{v} \mathbf{k} } \vert \, \widehat{e} \cdot \nabla_{\mathbf{k}} \widehat{H}_{\mathbf{k}} \, \vert \psi_{ n_{c} \mathbf{k} } \rangle \, .
    \label{eq:momentum_matrix_element}
\end{equation}

\noindent
where $\widehat{e}$ is a unit vector denoting the polarisation of the emitted or absorbed photon, with $\widehat{e} \cdot \nabla_{\mathbf{k}} \widehat{H}$ then denoting the directional derivative of the supercell Hamiltonian with respect to the component of $\mathbf{k}$ parallel to the photon polarisation. Evaluation of Eq.~\eqref{eq:momentum_matrix_element} requires calculation of the reciprocal space matrix representation of $\widehat{e} \cdot \nabla_{\mathbf{k}} \widehat{H}_{\mathbf{k}}$, which can be evaluated analytically by replacing $\widehat{H}_{b'b,\mathbf{k}} (z)$ by $\widehat{e} \cdot \nabla_{\mathbf{k}} \widehat{H}_{b'b,\mathbf{k}} (z)$ in Eq.~\eqref{eq:plane_wave_hamiltonian}.


The polarisation-averaged (squared) momentum matrix element is

\begin{equation}
    \vert \widetilde{p}_{ n_{c} n_{v} \mathbf{k} } \vert^{2} = \frac{1}{3} \left( \vert p_{ n_{c} n_{v} \mathbf{k} }^{ ( x ) } \vert^{2} + \vert p_{ n_{c} n_{v} \mathbf{k} }^{ ( y ) } \vert^{2} + \vert p_{ n_{c} n_{v} \mathbf{k} }^{ ( z ) } \vert^{2} \right) \, ,
    \label{eq:momentum_matrix_element_averaged}
\end{equation}

\noindent
which includes contributions associated with the emission of transverse electric (TE-) and transverse magnetic- (TM-) polarised photons, which respectively have polarisation in the plane perpendicular to the growth direction ($\widehat{e} = \widehat{x}$, $\widehat{y}$) and parallel to the growth direction ($\widehat{e} = \widehat{z}$). The SE rate is then evaluated in the quasi-equilibrium approximation as \cite{Lasher_PR_1964,Wu_JAP_2010,Fan_JAP_1996}

\begin{widetext} 
    \begin{equation}
    r_{\scalebox{0.7}{\text{sp}}} ( \hbar \omega ) = \frac{ e^{2} n_{r} \hbar \omega }{ \pi \epsilon_{0} m_{0}^{2} \hbar^{2} c^{3} } \sum_{ n_{c}, n_{v} } \left( \frac{ 2 \pi }{ L } \right)^{-1} \int_{ - \frac{ \pi }{L} }^{ + \frac{ \pi }{L} } \text{d} q \int \frac{ \text{d} \mathbf{k}_{\parallel} }{ ( 2 \pi )^{2} } \vert \widetilde{p}_{n_{c} n_{v} \mathbf{k} } \vert^{2} \, f_{e} ( E_{ n_{c} \mathbf{k} }, F_{e} ) \, \left[ 1 - f_{h} ( E_{ n_{v} \mathbf{k} }, F_{h} ) \right] \, \delta \left( E_{ n_{c} \mathbf{k} } - E_{ n_{v} \mathbf{k} } - \hbar \omega \right) \, ,
    \label{eq:spontaneous_emission_spectrum}
    \end{equation}
\end{widetext}

\noindent
where $E_{ n_{c} ( n_{v} ) \mathbf{k} }$ is the dispersion of conduction subband $n_{c}$ (valence subband $n_{v}$), $f_{ e (h) }$ is the electron (hole) Fermi-Dirac distribution function at temperature $T$, $F_{ e (h) }$ is the associated electron (hole) quasi-Fermi level corresponding to an injected electron (hole) carrier density $n$ ($p$), and the Dirac distribution $\delta$ imposes conservation of energy for photon emission.


Integration of $r_{\scalebox{0.7}{\text{sp}}} ( \hbar \omega )$ with respect to $\hbar \omega$ yields the radiative current density $J_{\text{rad}}$, which is given in the Boltzmann approximation by $J_{\text{rad}} = e B n p$, where $B$ is the radiative recombination coefficient. \cite{Wu_JAP_2010,Lasher_PR_1964,Fan_JAP_1996} $B$ can therefore be computed as

%
%

\begin{equation}
    B = \frac{ 1 }{ n p } \int r_{\scalebox{0.7}{\text{sp}}} ( \hbar \omega ) \, \text{d} ( \hbar \omega ) \, .
    \label{eq:radiative_recombination_coefficient}
\end{equation}

We impose net charge neutrality in our calculations -- i.e.~$n = p$ -- and fix $n = 10^{17}$ cm$^{-3}$ as a carrier density representative of LED operation. \cite{Repiso_JPDAP_2019} Numerical evaluation of Eq.~\eqref{eq:spontaneous_emission_spectrum} requires that the Dirac distribution be replaced by a finite-width lineshape. In our calculations we employ a hyperbolic secant lineshape, \cite{Marko_SR_2016} the spectral width (= 4 meV) of which was chosen based on qualitative comparison between our calculated SE spectra and the measured electroluminescence (EL) spectra of Ref.~\onlinecite{Zhou_APL_2019}. While this choice of lineshape function and spectral broadening impacts the calculated SE spectra, we note that it does not impact our calculated $B$ coefficients: integration with respect to $\hbar \omega$ in Eq.~\eqref{eq:radiative_recombination_coefficient} covers the entire range of possible photon energies, such that integration over any normalised lineshape function produces $B$ equal to that obtained assuming a Dirac lineshape.


\subsection{Technical considerations: choice of material parameters and mitigation of spurious solutions}
\label{sec:theory_technical_considerations}

The combination of physical transparency and computational efficiency offered by the $\mathbf{k} \cdot \mathbf{p}$ method have established it as the de facto standard approach to compute the electronic properties of III-V quantum-confined heterostructures. However, the conventional 8-band $\mathbf{k} \cdot \mathbf{p}$ Hamiltonian for zinc blende is known to provide an incomplete description of the electronic structure of tetrahedrally-bonded semiconductors, which can lead to the introduction of spurious solutions in the bulk band dispersion. \cite{Godfrey_PRB_1996} These spurious solutions can manifest as states having energies lying within the band gap and/or possessing complex-valued wave vectors. When performing heterostructure calculations, the presence of such solutions must be carefully considered and mitigated in order to obtain reliable electronic properties.

We begin our analysis by using the InAs and GaSb bulk band parameters (including the Varshni parameters for the bulk band gap temperature dependence), and the InAs/GaSb VB offset (= 0.56 eV), recommended by Vurgaftman et al. \cite{Vurgaftman_JAP_2001} We note that the reduction of the band gap of a semiconductor with increasing temperature has two contributions: lattice thermal expansion and electron-phonon coupling. \cite{Schluter_PRB_1975,Allen_PRB_1981} The Varshni relation is parametrised via fitting to the experimentally measured temperature-dependent band gap, therefore implicitly containing both contributions. In principle, lattice thermal expansion could be considered by using temperature-dependent lattice parameters to compute the strain in each layer of the structure. However, as described above, we neglect strain our present analysis. We therefore restrict the temperature dependence of our band parameters to the bulk InAs and GaSb band gaps only. Test calculations verify that, since we keep the Kane parameter $E_{P}$ fixed, this reduction in bulk band gap with increasing temperature leads to a reduction in band edge effective mass in line with experimental measurements (cf.~Eq.~\eqref{eq:sc} and, e.g., Ref.~\onlinecite{Schneider_PB_1998}). As we will describe below, some of the bulk band parameters recommended in Ref.~\onlinecite{Vurgaftman_JAP_2001} require modification to mitigate the deleterious impact of unphysical spurious solutions on the calculated SL eigenstates. Following Ref.~\onlinecite{Li_PRB_2010} we neglect the impact of strain on the SL band structure, due to the small lattice mismatch between InAs and GaSb. This is consistent with the x-ray diffraction measurements of Ref.~\onlinecite{Zhou_APL_2019}, which demonstrated lattice mismatch $\sim 10^{-5}$ with respect to the InAs substrate for the InAs/GaSb SLs against whose experimental characterisation we will benchmark our calculations in Sec.~\ref{sec:results_benchmark}.

In the 8-band $\mathbf{k} \cdot \mathbf{p}$ Hamiltonian the CB dispersion is described in part by the parameter $s_{c}$, alternatively denoted by $A_{c}$ in the literature, \cite{Foreman_PRB_1997,Veprek_PRB_2007}

\begin{equation}
     s_{c} = \frac{ 1 }{ m_{c}^{\ast} } - \frac{ E_{P} }{ 3 } \left( \frac{ 2 }{ E_{g} } + \frac{ 1 }{ E_{g} + \Delta_{0} } \right) \, ,
     \label{eq:sc}
\end{equation}

\noindent
where the first and second terms respectively describe the free-electron and remote-band contributions to the CB edge effective mass $m_{c}^{\ast}$, where $E_{g}$ and $\Delta_{0}$ are the $\Gamma$-point band gap and VB spin-orbit splitting, and $E_{P}$ is the Kane parameter. This parameter appears in the Hamiltonian as $\frac{ \hbar^{2} k^{2} }{2 m_{0} } \, s_{c}$ and, since $s_{c} < 0$ for most conventional semiconductors, this can lead to unphysical downward bending of the CB dispersion into the band gap at large $k = \vert \mathbf{k} \vert$. This is of particular importance for the simulation of short-period SL structures using the PWEM. Since $G_{m} \propto L^{-1}$, the plane wave basis set can readily sample wave vectors corresponding to large real values of the bulk $k_{z}$, and can therefore pollute the heterostructure eigenstates by sampling spurious large-$k_{z}$ CB states. Additionally, InAs/GaSb SLs demonstrate type-II-like (spatially indirect) carrier confinement, so that the Bloch character of a given eigenstate can abruptly change from being, e.g., predominantly CB-like to predominantly VB-like across an InAs/GaSb interface. These factors mandate careful selection of the (i) band parameters, and (ii) plane wave basis set, in numerical calculations.

\begin{figure*}[t!]
	\includegraphics[width=1.00\textwidth]{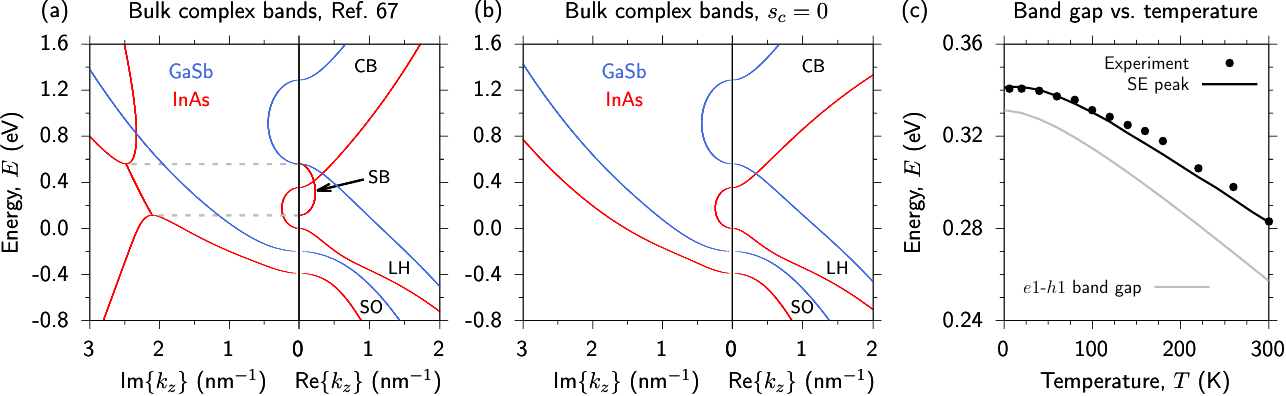}
	\caption{(a) Complex band structure of bulk InAs (solid red lines) and GaSb (solid blue lines) along [001], using an 8-band $\mathbf{k} \cdot \mathbf{p}$ Hamiltonian with the parameters of Ref.~\onlinecite{Vurgaftman_JAP_2001}. The arrow highlights the real $k_{z}$ component of a spurious band (SB) lying close in energy to the InAs CB edge, with the imaginary component of the SB lying in the same energy range in the left-hand panel. (b) As in (a), but using renormalised parameters in which $s_{c} = 0$. (c) Theory vs.~experiment comparison of the dependence of the emission energy vs.~temperature $T$ for an InAs/GaSb SL having equal InAs and GaSb layer thicknesses $t = 2.14$ nm (SL period $L = 2t = 4.28$ nm). The solid black line shows the calculated SE peak energy, obtained via a full evaluation of Eq.~\eqref{eq:spontaneous_emission_spectrum} for each temperature $T$ at injected electron and hole carrier densities $n = p = 10^{17}$ cm$^{-3}$. The solid grey line shows the calculated $e1$-$h1$ band gap at $\mathbf{k} = 0$. Experimental data (closed black circles) are from Ref.~\onlinecite{Zhou_APL_2019}.}
	\label{fig:complex_bands_and_gap_vs_temperature}
\end{figure*}

Solid red and blue lines in Fig.~\ref{fig:complex_bands_and_gap_vs_temperature}(a) respectively show the bulk complex band structure, at temperature $T = 300$ K, computed along $k_{z}$ for InAs and GaSb, with $\mathbf{k}_{\parallel} = 0$, using the 8-band $\mathbf{k} \cdot \mathbf{p}$ Hamiltonian parameterised following Ref.~\onlinecite{Vurgaftman_JAP_2001}. The zero of energy is set at the InAs VB maximum. The left- and right-hand panels of Fig.~\ref{fig:complex_bands_and_gap_vs_temperature}(a) respectively show the imaginary and real parts of $k_{z}$ associated with each bulk state at energy $E$. Note that we have omitted HH bands from Figs.~\ref{fig:complex_bands_and_gap_vs_temperature}(a) and~\ref{fig:complex_bands_and_gap_vs_temperature}(b) for simplicity, since the HH band decouples from the CB, LH and SO bands along [001] and is hence not relevant to the present discussion. Calculation of the bulk CB, LH and SO complex band dispersion -- described in detail in the Supplemental Information, which cites Refs.~\onlinecite{Kato_SM_2018}, \onlinecite{Cohen_PRB_1990}, \onlinecite{Bahder_PRB_1990}, \onlinecite{Foreman_PRB_1997} and \onlinecite{Chang_PRB_1982,Das_PRA_2022,Chen_PRB_1992,Schulman_PRB_1985,Chang_PRB_1985} -- is performed based on a reduced 3-band Hamiltonian that appears in the block-diagonalised 8-band Hamiltonian along [001] \cite{Cohen_PRB_1990,Bahder_PRB_1990}

\begin{widetext}
    \begin{equation}
	H_{3 \times 3} ( k_{z} ) = \left( \begin{array}{ccc}
            E_{g} + \frac{ \hbar^{2} }{ 2 m_{0} } \, s_{c} \, k_{z}^{2} & \sqrt{ \frac{2}{3} } P k_{z}                                                       & - \frac{1}{ \sqrt{3} } P k_{z} \\
             \sqrt{ \frac{2}{3} } P k_{z}                               & - \frac{ \hbar^{2} }{ 2 m_{0} } \left( \gamma_{1} + 2 \gamma_{2} \right) k_{z}^{2} & \sqrt{2} \frac{ \hbar^{2} }{ m_{0} } \gamma_{2} \, k_{z}^{2} \\
            - \frac{1}{ \sqrt{3} } P k_{z}                              & \sqrt{2} \frac{ \hbar^{2} }{ m_{0} } \gamma_{2} \, k_{z}^{2}                       & - \Delta_{0} - \frac{ \hbar^{2} }{ 2 m_{0} } \gamma_{1} \, k_{z}^{2} \\
        \end{array} \right) \, \begin{array}{l} \vert \frac{1}{2} ; \pm \frac{1}{2} \rangle \\ ~ \vspace{-0.15cm} \\ \vert \frac{3}{2} ; \pm \frac{1}{2} \rangle \\ ~\vspace{-0.15cm} \\ \vert \frac{1}{2} ; \pm \frac{1}{2} \rangle \\ \end{array} \, ,
	   \label{eq:3x3_hamiltonian}
    \end{equation}
\end{widetext}

\noindent
where $P$ ($= \sqrt{ \frac{ m_{0} E_{P} }{ 2 } }$, where $m_{0}$ is the free electron mass) is the inter-band (Kane) momentum matrix element, and $\gamma_{1}$ and $ \gamma_{2}$ are the modified VB Luttinger parameters. \cite{Cohen_PRB_1990,Bahder_PRB_1990}


The InAs complex band dispersion in Fig.~\ref{fig:complex_bands_and_gap_vs_temperature}(a) reveals the presence of a spurious band (SB), having complex-valued $k_{z}$ and which straddles the CB minimum in the energy range 0.11 -- 0.55 eV (highlighted by horizontal dashed grey lines). This SB originates due to the presence of a large negative value $s_{c} \approx -11.7$, occurring in the energy range in which the purely imaginary (evanescent) band having $\text{Im} \lbrace k_{z} \rbrace \approx 2.5$ nm$^{-1}$ across the full plotted energy range intersects the evanescent band originating from the SO VB. Large negative values of $s_{c}$ most often give rise to a SB at large real $k_{z}$, due to the aforementioned unphysical downward dispersion of the CB. Such unphysical CB dispersion is \textit{not} observed in the real band dispersion of the 8-band model parameterised following Ref.~\onlinecite{Vurgaftman_JAP_2001}. However, the combination of $s_{c}$ and the LH and SO effective masses of Ref.~\onlinecite{Vurgaftman_JAP_2001} (determined by the modified VB Luttinger parameters $\gamma_{1}$ and $\gamma_{2}$) produce a SB having complex-valued $k_{z}$ that is not visible in the computed real bulk band dispersion, but which is nonetheless capable of polluting the results of heterostructure calculations. Test calculations for InAs/GaSb SLs using this parameter set demonstrated the introduction of spurious solutions lying close in energy to the CB edge for plane wave basis sets containing as few as $2M + 1 = 7$ plane waves -- a consequence of the appearance of the SB at small real $k_{z}$ -- with these spurious solutions moving downwards in energy into the band gap, and subsequently into the VB, with increasing $M$.

Our combined analytical and numerical analysis of Eq.~\eqref{eq:3x3_hamiltonian} (cf.~Supplemental Information) demonstrates that, as $s_{c}$ approaches zero, this SB evolves from having complex-valued wave vector with small real $k_{z}$, towards being an evanescent state having purely imaginary $k_{z}$ with $\vert \text{Im} \lbrace k_{z} \rbrace \vert \to \infty$ as $s_{c} \to 0$. To mitigate the impact of this SB we follow the approach suggested by Foreman, \cite{Foreman_PRB_1997} and set $s_{c} = 0$. We achieve this by decreasing the Kane parameter $E_{P}$ to fix $s_{c} = 0$ in both InAs and GaSb, leaving the value of the bulk CB edge effective mass unchanged in each case. This parameter renormalisation is consistent with the analysis of Veprek et al., \cite{Veprek_PRB_2007} who demonstrated that reducing $E_{P}$ -- which, in turn, adjusts the modified VB Luttinger parameters in addition to $s_{c}$ -- is an effective approach to eliminate spurious solutions by enforcing ellipticity of the 8-band $\mathbf{k} \cdot \mathbf{p}$ Hamiltonian. Solid red and blue lines in Fig.~\ref{fig:complex_bands_and_gap_vs_temperature}(b) respectively show the calculated InAs and GaSb bulk complex band dispersion using a renormalised parameter set in which $s_{c} = 0$. The complex band dispersion of InAs close in energy to the band gap is now in very good agreement with that obtained from full-band atomistic tight-binding calculations. \cite{Chang_PRB_1985,Carrillo-Nunez_SSE_2015} We therefore retain these renormalised parameters for our analysis of the InAs/GaSb SL electronic and optical properties.

We note that while enforcing $s_{c} = 0$ removes a complex-valued SB from close to the zone centre in InAs, this SB is replaced by an evanescent state having large imaginary wave vector. Calculation of bound heterostructure eigenstates, which are formally equivalent to a linear combination of bulk states having complex-valued wave vectors, typically requires matching of envelope function amplitudes and (appropriately weighted) spatial derivatives at material interfaces. \cite{Chen_PRB_1992} Given that our choice of renormalised band parameters replaces a SB by an evanescent state in InAs, we might then expect that this will impact the character of the computed eigenstates. As described by Foreman in Ref.~\onlinecite{Foreman_PRB_1997}, forcing $s_{c} = 0$ removes the requirement that the CB Bloch component of the envelope function be continuous across an interface (cf.~Supplemental Information). We will observe, and discuss, below that this is indeed the case for our calculated InAs/GaSb SL bound electron states.


\section{Results}
\label{sec:results}


\subsection{Benchmark calculations: SL emission energy and its temperature dependence}
\label{sec:results_benchmark}

Motivated by the structures considered in Ref.~\onlinecite{Zhou_APL_2019}, we begin our analysis by considering InAs/GaSb SLs having equal InAs and GaSb layer thicknesses $t_{\text{InAs}} = t_{\text{GaSb}} \equiv t$. The corresponding length of our calculational supercell -- i.e.~the SL period -- is $L = 2t$. Firstly, in order to establish an appropriate choice of plane wave basis set, we take exemplar structures and compute the $\mathbf{k} = 0$ bound state energies as a function of the number of plane waves -- i.e.~by varying $M$ in Eq.~\eqref{eq:plane_wave_expansion}. A basis set containing $2M + 1 = 51$ plane waves was found to be sufficient to converge all calculated bound state energies to within 0.1 meV, with respect to further increases in the size of the basis set. We therefore set $M = 25$ for all of our calculations. This choice of plane wave basis set generates a distinct plane wave Hamiltonian matrix $\widetilde{H}_{\mathbf{k}}$ of size $8 ( 2M + 1 ) \times 8 ( 2M + 1 ) = 408 \times 408$ at each wave vector $\mathbf{k}$.

Having established an appropriate choice of plane wave basis set we proceed by performing benchmark calculations of the SL peak energy and its temperature dependence, and compare to the EL measurements of Ref.~\onlinecite{Zhou_APL_2019}. Here, it is important to note that the measured EL peak energy does not correspond directly to the SL band gap -- between the lowest energy bound electron state $e1$ and the highest energy bound hole state $h1$ at $\mathbf{k} = 0$ -- due to band filling effects under carrier injection. Band filling effects act to blueshift the EL/SE peak relative to the SL band gap, such that both the in-plane ($\mathbf{k}_{\parallel}$) and miniband ($q$) dispersion play important roles in determining not only the EL/SE peak energy, but also its temperature dependence.

The results of our benchmark calculations are summarised in Fig.~\ref{fig:complex_bands_and_gap_vs_temperature}(c), where closed black circles depict the EL data of Ref.~\onlinecite{Zhou_APL_2019}. We present two sets of theoretical data in Fig.~\ref{fig:complex_bands_and_gap_vs_temperature}(c). Firstly, we compute the emission energy as the peak energy of the full SL SE spectrum computed via Eq.~\eqref{eq:spontaneous_emission_spectrum}. This peak energy is expected to correspond closely to the experimentally measured EL peak energy. We therefore proceed by taking the measured $T = 300$ K EL peak energy ($= 0.283$ eV $= 4.38$ $\mu$m) as a reference, and adjust the layer thickness $t$ to match our calculated $T = 300$ K SE peak energy to this reference. This yields $t = 2.14$ nm ($L = 4.28$ nm), which corresponds well to the nominal $t = 2.3$ nm of Ref.~\onlinecite{Zhou_APL_2019}. This minor discrepancy between the nominal and best-fit layer thicknesses in our calculations vs.~the device structures investigated in Ref.~\onlinecite{Zhou_APL_2019} is expected due to (i) the incorporation of a small fraction ($\lesssim 10$\%) of As in the GaSb layers, to minimise lattice mismatch with respect to InAs, and (ii) the fact that the SL regions of the devices were lightly n-doped, which is not considered in our calculations. Keeping this value of $t$ fixed for all of our benchmark calculations we compute the $T$-dependent SE peak energy, shown by the solid black line in Fig.~\ref{fig:complex_bands_and_gap_vs_temperature}(c). We note good quantitative agreement with the measurements of Ref.~\onlinecite{Zhou_APL_2019} (closed black circles). Secondly, the solid grey line in Fig.~\ref{fig:complex_bands_and_gap_vs_temperature}(c) shows the calculated $T$-dependent $\mathbf{k} = 0$ $e1$-$h1$ SL band gap for the same structure. Due to the Burstein-Moss effect, this fundamental SL band gap is lower in energy than the peak energies obtained from the SE calculations. To quantify the $T$-dependence of these two sets of calculated energies for InAs/GaSb SLs, we fit to the empirical Varshni relation \cite{Varshni_physica_1967,Vurgaftman_JAP_2001}

\begin{equation}
     E_{\text{gap/peak}} (T) = E_{\text{gap/peak}} (0) - \frac{ \alpha \, T^{2} }{ T + \beta } \, ,
     \label{eq:varshni_fit}
\end{equation}

\noindent
where $E_{\text{gap/peak}} (0)$ is the zero-temperature SL band gap or SE peak energy.

To facilitate comparison to literature experimental data, we follow Ref.~\onlinecite{Klein_JPDAP_2011} by fixing $\beta = 270$ K when fitting the theoretical data of Fig.~\ref{fig:complex_bands_and_gap_vs_temperature}(c) via Eq.~\eqref{eq:varshni_fit}. We recall that our analysis employs only the bulk InAs and GaSb Varshni parameters as input to the SL electronic structure calculations, with the subsequent Varshni fits to the temperature-dependent $e1$-$h1$ band gap and SE peak energies then being predictions of the full SL calculations. The resultant best-fit values of the Varshni parameter $\alpha$ are 0.383 and 0.493 meV K$^{-1}$ for the calculated full SE peak energy (solid black line) and $\mathbf{k} = 0$ SL band gap (solid grey line), respectively. We note that $\alpha = 0.383$ meV K$^{-1}$ for the full SE calculation -- i.e.~including the full $\mathbf{k}$-dependent in-plane and miniband dispersion, and optical matrix elements -- is in close quantitative agreement with the range $\alpha = 0.30$ -- 0.37 meV K$^{-1}$ of experimental values of Ref.~\onlinecite{Klein_JPDAP_2011}.



\subsection{Equal layer thickness: SL electronic structure and its impact on radiative recombination}
\label{sec:results_equal_thickness}

Despite that the $T$-dependence of the bulk InAs and GaSb band gaps used as input to our calculations -- by applying Eq.~\eqref{eq:varshni_fit} in conjunction with the parameters of Ref.~\onlinecite{Vurgaftman_JAP_2001} -- are treated identically for both sets of calculations shown in Fig.~\ref{fig:complex_bands_and_gap_vs_temperature}(c), we note that the resultant best-fit values of $\alpha$ vary by close to 30\%. This describes that, relative to the SL band gap, filling of in-plane bands and minibands by injected carriers acts not only to blueshift the SE peak energy at fixed $T$, but also to decrease its $T$ dependence compared to that of the bulk band gaps of the SL's constituent materials. Having demonstrated the ability of our model to quantitatively capture the experimentally measured $T$-dependence of the SL emission energy, we now turn our attention to a detailed interpretation of radiative recombination in this equal layer thickness SL ($t = 2.14$ nm, $L = 2t$) based on its calculated electronic structure. Unless otherwise stated, all calculations below are performed at $T = 300$ K.


\begin{figure*}[t!]
	\includegraphics[width=1.00\textwidth]{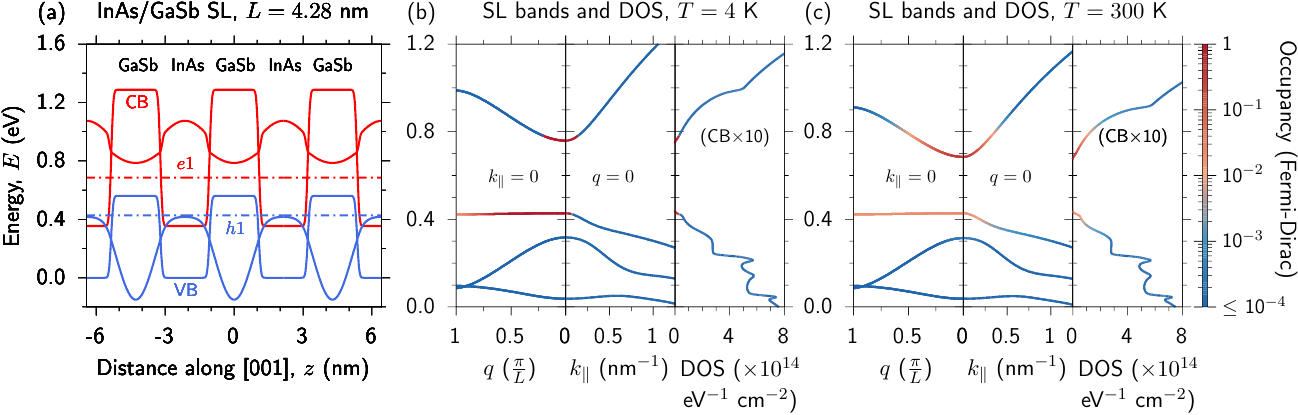}
	\caption{(a) Calculated CB (solid red line) and VB (solid blue line) offsets, illustrated for three periods of a InAs/GaSb SL having equal InAs and GaSb layer thickness $t = 2.14$ nm (SL period $L = 4.28$ nm). The dash-dotted red (blue) line denotes the energy of the lowest energy bound electron state $e1$ (highest energy bound hole state $h1$). The $e1$ ($h1$) probability density is shown using a solid red (blue) line, shifted so that the zero of probability density lies at the $e1$ ($h1$) energy. (b) and (c) Calculated band structure and DOS at temperature (b) $T = 4$ K, and (c) $T = 300$ K, for the equal layer thickness SL of (a). The left-hand, centre and right-hand panels in (b) and (c) respectively show the calculated miniband dispersion, in-plane band dispersion and DOS. The CB DOS is scaled $\times 10$ to enhance visibility. The line colour is weighted by the Fermi-Dirac occupancy, for injected electron and hole carrier densities $n = p = 10^{17}$ cm$^{-3}$.}
	\label{fig:equal_thickness_electronic}
\end{figure*}


\paragraph*{Carrier localisation --}

We begin by considering the electron and hole localisation, summarised in Fig.~\ref{fig:equal_thickness_electronic}(a). Solid red and blue lines respectively show the computed CB and VB offsets (demonstrating the type-III band alignment), and also the probability densities associated respectively with the $e1$ and $h1$ $\mathbf{k} = 0$ bound states. Dash-dotted red and blue lines respectively denote the computed $e1$ and $h1$ bound state energies at $\mathbf{k} = 0$, which in each case serves as the zero of the plotted probability density. We note that the probability density for each state ($n = e1$, $h1$) is computed at each position $z$ by summing over $\vert F_{nb\mathbf{k}} (z) \vert^{2}$ with respect to the bulk Bloch band index $b$ at $\mathbf{k} = 0$. The $e1$ probability density therefore contains CB, LH and SO -- i.e.~$\vert J; m_{J} \rangle = \vert \frac{1}{2}; \pm \frac{1}{2} \rangle$, $\vert \frac{3}{2}; \pm \frac{1}{2} \rangle$ and $\vert \frac{1}{2}; \pm \frac{1}{2} \rangle$ -- contributions, while the $h1$ probability density contains only HH -- i.e.~$\vert J; m_{J} \rangle = \vert \frac{3}{2}; \pm \frac{3}{2} \rangle$ -- contributions. While our electronic structure calculations are performed for supercells consisting of a single SL period we have, for clarity, exploited the periodic boundary conditions to plot the band offsets and probability densities in Fig.~\ref{fig:equal_thickness_electronic}(a) over three SL periods.

We note that the $h1$ state is tightly confined within GaSb layers in the structure, with its probability density dropping rapidly to near-zero values at the centre of InAs layers as a consequence of the large VB offset and HH VB edge effective mass in GaSb. Conversely, we note that the $e1$ state is, while primarily localised in InAs layers, comparatively delocalised. The $e1$ probability density retains significant non-zero values throughout GaSb layers. This partial delocalisation of the $e1$ state is a consequence of the combination of low InAs CB edge effective mass and low GaSb layer thickness, and reflects strong coupling between electrons in neighbouring SL periods. As we will describe below, this leads to (i) strong dispersion of the $e1$ miniband along $q$, and (ii) significant electron-hole spatial overlap, leading to optical matrix elements that can be considered high for a structure possessing spatially indirect electron and hole confinement.

Before proceeding, we note a technical consideration relevant to the calculation of the carrier envelope functions. Our preliminary calculations of the $e1$ probability density demonstrated oscillations, \cite{Murphy_NUSOD_2021} suggesting the potential presence of a spurious contribution to the probability density, despite the choice of renormalised parameters having $s_{c} = 0$. Detailed analysis confirms that these oscillations are \textit{not} a consequence of a spurious contribution to the $e1$ eigenstate, but rather are a consequence of the use of a plane wave basis set in conjunction with $s_{c} = 0$. \cite{Foreman_PRB_1997} In this case, oscillations in the $e1$ probability density arise in the presence of spatially abrupt InAs/GaSb interface, due to an accompanying discontinuity in the $e1$ envelope function (cf.~Supplemental Information). Mathematically, these oscillations arise due to the description of the envelope function discontinuity in the PWEM -- i.e.~via a truncated Fourier series (cf.~Eq.~\eqref{eq:plane_wave_expansion}) -- as a manifestation of the Gibbs phenomenon. \cite{Stone_Cambridge_2012} To eliminate this behaviour we have softened the InAs/GaSb interfaces by convoluting the initial piecewise continuous real space band offset profile with a Gaussian distribution of standard deviation $\sigma = 7.5 \times 10^{-2}$ nm. This generates the continuous band offset profile shown in Fig.~\ref{fig:equal_thickness_electronic}(a), while still allowing for analytical evaluation of the matrix elements of $\widetilde{H}_{\mathbf{k}}$ via the convolution theorem. Even with this applied InAs/GaSb intermixing we observe in Fig.~\ref{fig:equal_thickness_electronic}(a) that the magnitude of the $e1$ probability density changes abruptly across InAs/GaSb interfaces. This is expected based on Foreman's analysis of the behaviour of a CB envelope function when $s_{c} = 0$, \cite{Foreman_PRB_1997} and is also consistent with recent atomistic calculations of the $e1$ probability density based on the tight-binding method. \cite{Kato_SM_2018}


\paragraph*{SL band structure --}

Next, we turn our attention to the calculated SL band structure and DOS, shown in Figs.~\ref{fig:equal_thickness_electronic}(b) and~\ref{fig:equal_thickness_electronic}(c) for $T = 4$ and 300 K respectively. In Figs.~\ref{fig:equal_thickness_electronic}(b) and~\ref{fig:equal_thickness_electronic}(c) the left-hand and centre panels respectively show the calculated miniband and in-plane band dispersion, while the right-hand panel shows the calculated DOS. In each case the calculated CB DOS, which is significantly lower than the VB DOS, has been multiplied by 10 for improved visibility. Line colours denote the Fermi-Dirac band occupancy, which ranges from 1 (red) to $\leq 10^{-4}$ (blue), corresponding to the electron and hole quasi-Fermi levels computed for injected carrier densities $n = p = 10^{17}$ cm$^{-3}$. We note that there is minimal qualitative difference between the calculated $T = 4$ and 300 K SL band structures, with the only significant quantitative difference being the 74 meV reduction in SL band gap between $T = 4$ and 300 K (cf.~Fig.~\ref{fig:complex_bands_and_gap_vs_temperature}(c)). Examining the computed miniband structures we firstly note that, as described above, the partial $e1$ delocalisation is reflected by the presence of highly dispersive $e1$ minibands. As a consequence, electrons only occupy states over a limited range of $q$. At low temperature $T = 4$ K, this range is strongly limited by the step-like nature of the Fermi-Dirac distribution function, such that electrons only occupy states within the first $\approx 20$\% of the SL BZ -- i.e.~$\vert q \vert \lesssim \frac{ \pi }{ 5L }$. Conversely, the strong $h1$ localisation suppresses electronic coupling of VB edge hole states in neighbouring SL periods, which is reflected by nearly dispersionless $h1$ minibands. Holes therefore occupy $h1$ miniband states throughout the entirety of the SL BZ, even at low $T$.

This mismatch in the dispersion of the $e1$ and $h1$ minibands, combined with strict $\mathbf{k}$-selection for optical transitions, acts to reduce the radiative recombination rate. Specifically, holes occupying $h1$ miniband states at large $\vert q \vert$ are without electrons having matching wave vector, and hence have no available electrons with which to recombine. We therefore expect, and will demonstrate below, that the presence of this mismatched $e1$ and $h1$ miniband dispersion leads to a reduction of $B$ at fixed temperature and carrier density vs.~an equivalent QW structure (i.e.~without miniband dispersion). We note that this behaviour mirrors that associated with the mismatch of the in-plane CB and VB edge DOS in conventional QW structures, which served as a key motivation for the development of strain-engineered QW structures to improve the performance of LEDs and diode lasers. While the mismatch in the range of $q$ over which electrons and holes occupy miniband states is exacerbated at low temperature, we note that this behaviour is partially mitigated at $T = 300$ K. As $T$ is increased to 300 K the emergence of a high-energy tail in the electron Fermi-Dirac distribution above the quasi-Fermi level allows electrons to occupy $e1$ miniband states over a significantly larger fraction of the SL BZ (cf.~Fig.~\ref{fig:equal_thickness_electronic}(c) vs.~Fig.~\ref{fig:equal_thickness_electronic}(b)), thereby providing a radiative ($\Delta \mathbf{k} = 0$) recombination pathway for holes having larger $\vert q \vert$. The impact of this behaviour on the magnitude of $B$ and its $T$ dependence will be quantified below.


\begin{figure*}[t!]
	\includegraphics[width=1.00\textwidth]{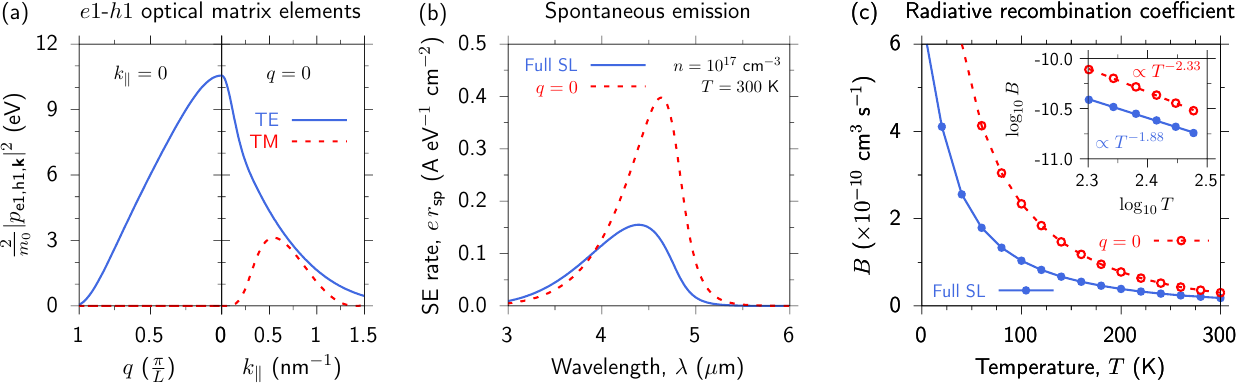}
	\caption{(a) Calculated TE-polarised (solid blue line) and TM-polarised (dashed red line) squared optical (momentum) matrix elements between the $e1$ and $h1$ subbands, as a function of the SL wave vector $q$ (left-hand panel) and in-plane wave vector $k_{\parallel}$ (right-hand panel). (b) Comparison of SE spectra for an InAs/GaSb SL having equal InAs and GaSb layer thicknesses $t = 2.14$ nm (SL period $L = 2t = 4.28$ nm), calculated at temperature $T = 300$ K and injected carrier density $n = 10^{17}$ cm$^{-3}$, directly using Eq.~\eqref{eq:spontaneous_emission_spectrum} (solid blue line) and via a simplified calculation neglecting miniband dispersion by fixing $q = 0$ in Eq.~\eqref{eq:plane_wave_expansion} (dashed red line). (c) Comparison of the $T$-dependent radiative recombination coefficient $B$ obtained from the $q = 0$ (open red circles) and full SE calculations (closed blue circles). Inset: $\log_{10} B$ vs.~$\log_{10} T$ for $T = 200$ -- 300 K, where $B \propto T^{s}$ with $s$ equal to the log-log slope.}
	\label{fig:equal_thickness_optical}
\end{figure*}


\paragraph*{Optical matrix elements --}

Using the SL Hamiltonian $\widetilde{H}_{\mathbf{k}}$ and eigenstates $\vert \psi_{ n \mathbf{k} } \rangle$ we explicitly compute the TE- and TM-polarised optical matrix elements for $e1$-$h1$ transitions using Eq.~\eqref{eq:momentum_matrix_element}. The results of these calculations are summarised in Fig.~\ref{fig:equal_thickness_optical}(a), where the left- and right-hand panels respectively show the optical matrix elements vs.~$q$ for $k_{\parallel} = 0$, and vs.~$k_{\parallel}$ for $q = 0$. TE- and TM-polarised matrix elements are respectively denoted by solid blue and dashed red lines. We present the squared of the optical (momentum) matrix elements as $\frac{ 2 }{ m_{0} } \vert p_{ n_{c} n_{v} \mathbf{k} }^{ ( e ) } \vert^{2}$ -- i.e.~having units of energy, so that a natural reference is provided by the bulk Kane parameters $E_{P}$. The in-plane matrix elements obey conventional selection rules. At $k_{\parallel} = 0$ the $h1$ state is purely HH-like, possessing no $p_{z}$-like orbital component, so that the TM-polarised transition is symmetry forbidden. Away from $k_{\parallel} = 0$ the $h1$ band begins to hybridise with other bands and acquire LH-like Bloch character -- and, to a lesser extent, SO- and CB-like character -- thereby acquiring a $p_{z}$-like orbital component such that the TM-polarised matrix element initially grows with increasing $k_{\parallel}$ before again reducing as $k_{\parallel}$ increases further (due in part to the growth in the $E_{e1,k_{\parallel}} - E_{h1,k_{\parallel}}$ energy gap). The TE-polarised matrix element is non-zero across the full range of $k_{\parallel}$ across which $e1$ and $h1$ band states are occupied reflecting that, despite strong band hybridisation away from $k_{\parallel} = 0$, the $e1$ and $h1$ states respectively retain significant CB- and HH-like Bloch character. We note that the calculated TE-polarised optical matrix element at $\mathbf{k} = 0$ achieves a value that is 64.0\% of the renormalised InAs Kane parameter $E_{P} = 16.49$ eV employed in the electronic structure calculation. This substantial value is a consequence of the partial $e1$ delocalisation described above, which allows for large $e1$-$h1$ spatial overlap despite the nominally type-II-like carrier confinement.

At $k_{\parallel} = 0$ the TE-polarised matrix element decreases with increasing $\vert q \vert$, becoming zero at $\vert q \vert = \frac{ \pi }{ L }$, as has been observed in some previous calculations (see, e.g., Fig.~6(a) of Ref.~\onlinecite{Li_PRB_2010}). This behaviour traces its origin to an interplay between the evolution of the $e1$ and $h1$ envelope functions with increasing $\vert q \vert$, combined with the spatially indirect carrier confinement in the SL. As a consequence of the spatially indirect carrier confinement, the peaks of the $e1$ and $h1$ probability densities in this exemplar equal layer thickness structure are offset by a distance $\Delta z = t = \frac{L}{2}$. As such, traversing the $e1$ and $h1$ minibands -- i.e.~varying $q$ from 0 to $\frac{ \pi }{ L }$ while maintaining $k_{\parallel} = 0$ -- imparts a phase factor of $\exp( i q \Delta z ) = \exp( i \frac{ q L}{2} )$ between the $e1$ and $h1$ envelope functions. This phase difference attains a value of $\frac{ \pi }{ 2 }$ at $q = \frac{ \pi }{ L }$, with the corresponding phase factor $\exp( i \frac{ \pi }{ 2 } ) = i$ then being tantamount to introducing opposite parity between the $e1$ and $h1$ envelope functions (vs.~their having equal parity at $\mathbf{k} = 0$; cf.~Supplemental Information). As a result, the TE-polarised optical matrix element between the $e1$ and $h1$ states vanishes at $\vert q \vert = \frac{ \pi }{ L }$. We note that this $q = \pm \frac{ \pi }{ L }$ ``selection rule'' is a consequence of neglecting MIA in continuum calculations (cf.~Sec.~\ref{sec:theory_electronic}). Inclusion of MIA -- either implicitly in atomistic calculations, or via an empirical potential in continuum calculations -- lowers the symmetry of the eigenstates, so that the $e1$-$h1$ transition attains a much reduced but formally non-zero matrix element at $\vert q \vert = \frac{ \pi }{ L }$. \cite{Chang_PRB_1985,Li_PRB_2010} From the perspective of the present analysis we emphasise that electrons do not occupy $e1$ miniband states at large $\vert q \vert$, so that this slight underestimation of the TE-polarised optical matrix element at large $\vert q \vert$ will not impact the calculated SE spectra or values of $B$ derived therefrom. We find the calculated optical matrix elements of Fig.~\ref{fig:equal_thickness_optical}(a) to be in good overall agreement with previous work.


\paragraph*{SE spectra and $B$ coefficient --}

Using the computed SL electronic structure and optical matrix elements for this exemplar equal layer thickness InAs/GaSb SL, we evaluate the SE spectrum for injected carrier density $n = p = 10^{17}$ cm$^{-3}$ via Eq.~\eqref{eq:spontaneous_emission_spectrum}. For comparative purposes, we show the results of two calculations in Fig.~\ref{fig:equal_thickness_optical}(b). The solid blue line shows the full calculated SL SE spectrum, which explicitly accounts for the $q$ dependence of the SL electronic structure and optical matrix elements. The dashed red line shows the results of a simplified calculation in which we fix $q = 0$ and neglect $q$ dependence -- i.e.~treating a single SL period as a conventional QW by omitting miniband dispersion. We recall that the $T$ dependence of the ``full SL'' SE peak energy is shown by the solid black line in Fig.~\ref{fig:complex_bands_and_gap_vs_temperature}(c). Comparison of these two sets of SE calculations elucidates the impact of SL miniband formation on radiative recombination. Firstly, the strong dispersion of the $e1$ miniband in a short-period SL results in electrons occupying CB states at higher energy (cf.~Fig.~\ref{fig:equal_thickness_electronic}(b) and~\ref{fig:equal_thickness_electronic}(c)), leading in turn to a blueshift of the SL peak energy compared to an equivalent isolated QW. Secondly, the aforementioned mismatch in the occupancy of $e1$ and $h1$ minibands at large values of $\vert q \vert$ reduces the number of electron-hole pairs that can undergo radiative recombination, thereby reducing the peak SE rate. This reduction of the radiative recombination rate in response to miniband formation highlights the requirement for careful band structure engineering -- potentially exploiting alloying and strain in addition to adjusting the thickness of electron- and hole-confining layers -- to optimise the performance of SL-based emitters.

For $n = p = 10^{17}$ cm$^{-3}$ we compute at $T = 300$ K that 100\% of injected electrons occupy $e1$ band states, while $> 99.6$\% of injected holes occupy $h1$ band states. The SE spectrum and $B$ are therefore determined almost entirely by radiative recombination of $e1$ electrons and $h1$ holes. For our high-throughput calculations of $B$ below, we therefore compute SE spectra by restricting the evaluation of Eq.~\eqref{eq:spontaneous_emission_spectrum} to only the $n_{c} = e1$ conduction and $n_{v} = h1$ valence subbands. Further test calculations reveal that the SE is dominated by TE-polarised $e1$-$h1$ transitions. For example, integrating over the $T = 300$ K SE spectrum obtained from a calculation that omits the TM-polarised contribution to $\vert \widetilde{p}_{ n_{c} n_{v} \mathbf{k} } \vert^{2}$ (cf.~Eq.~\eqref{eq:momentum_matrix_element_averaged}) yields a radiative recombination coefficient $B$ that is 96.4\% of that obtained from the full calculation including both TE- and TM-polarised transitions. That this is the case is expected based on our calculated band occupancies and optical matrix elements: at $T = 300$ K the $e1$ and $h1$ bands only have significant occupancies $\gtrsim 10^{-2}$ for in-plane wave vectors $k_{\parallel} \lesssim 0.25$ nm$^{-1}$ (cf.~Fig.~\ref{fig:equal_thickness_electronic}(b)), in which range the TE-polarised $e1$-$h1$ optical matrix element significantly exceeds its TM-polarised counterpart (cf.~Fig.~\ref{fig:equal_thickness_optical}(a)).

The results of our calculations of the $T$-dependent $B$ coefficient for this equal layer thickness SL structure are summarised in Fig.~\ref{fig:equal_thickness_optical}(c). Closed blue circles denote the $B$ coefficients obtained from full SL calculations, while open red circles summarise the $B$ coefficients obtained from simplified $q = 0$ calculations. In line with our calculated peak SE rates, we firstly note that the SL demonstrates reduced $B$ coefficient at fixed temperature compared to an equivalent QW structure. We secondly note that the degree to which the SL $B$ coefficient is reduced in the SL is strongly dependent on temperature. Miniband formation drives strong reduction of $B$ at low $T$, with the calculated SL radiative recombination coefficient $B = 6.16 \times 10^{-10}$ cm$^{3}$ s$^{-1}$ at $T = 4$ K being only 32.5\% of that obtained from the equivalent $q = 0$ calculation. This disparity is strongly reduced at $T = 300$ K, with $B = 1.81 \times 10^{-11}$ cm$^{3}$ s$^{-1}$ being 60.0\% of that obtained from the equivalent ``$q = 0$'' calculation. This behaviour can be understood in light of the $T$-dependent $e1$ miniband occupancy (cf.~Fig.~\ref{fig:equal_thickness_electronic}(b) and~\ref{fig:equal_thickness_electronic}(c)). In a conventional QW, the large difference in CB and VB edge effective masses leads to electrons (holes) occupying $e1$ ($h1$) subband states over a larger range of energy (in-plane wave vector $\mathbf{k}_{\parallel}$) than holes (electrons). \cite{Reilly_IEEEJQE_1994,Sweeney_JAP_2019} At fixed carrier density this mismatch in occupancy is exacerbated with increasing $T$, as the electron (hole) quasi-Fermi level moves upwards (downwards) in energy, such that the $B$ coefficient of a QW decreases strongly with increasing $T$. In the SL structures under consideration, the mismatch in $e1$ and $h1$ subband occupancy at fixed $T$ is exacerbated by miniband formation. However, the reduction in the $e1$ and $h1$ occupancy mismatch along $q$ with increasing $T$ constitutes a competing effect, acting to mitigate the degree to which the number of electron-hole pairs available to undergo radiative recombination is reduced. We therefore expect that the $T$-dependence of $B$ should be reduced in the SL compared to the equivalent QW.

The $T$ dependence of $B$ is summarised in the inset to Fig.~\ref{fig:equal_thickness_optical}(c), where closed blue (open red) circles display a log-log plot of $B$ vs.~$T$ in the range $T = 200$ -- 300 K for the full SL ($q = 0$) calculations. Here, we take $B \propto T^{s}$ and obtain $s$ as the slope of a linear fit -- shown using a solid blue or dashed red line -- to the log-log data. In the standard analytical treatment of $B$ -- which makes the simplifying assumptions of CB and VB parabolicity, temperature-independent band gap, and non-degenerate (Maxwell-Boltzmann) carrier statistics -- this results in $B \propto E_{g} \, T^{-D/2}$ in a semiconductor having band gap $E_{g}$ and $D$-dimensional band dispersion. \cite{Lasher_PR_1964,Haug_APB_1987} For our exemplar SL ($D = 3$) we verify this expected behaviour by performing a test calculation in which we omit the $T$ dependence of the InAs and GaSb band gaps, and reduce the carrier density to $10^{15}$ cm$^{-3}$ so that the electron and hole Fermi-Dirac distribution functions mimic the non-degenerate (Boltzmann) regime. The $T$ dependence of $B$ then originates solely from that of the carrier distribution functions (cf.~Eq.~\eqref{eq:spontaneous_emission_spectrum}). This test calculation yields $s = -1.49$, in line with the expected analytical result $B \propto T^{-3/2}$. The enhanced temperature sensitivity $B \propto T^{-1.88}$ obtained from the full $T$-dependent SL calculation is therefore attributed to the reduction of the SL band gap with increasing $T$ (cf.~Fig.~\ref{fig:complex_bands_and_gap_vs_temperature}(c)). The equivalent $q = 0$ calculation demonstrates higher sensitivity of $B$ on $T$, with $s = -2.33$, confirming our expectation that miniband formation can act to reduce the dependence of $B$ on $T$ in a SL in which there exists a significant mismatch in the dispersion of the $e1$ and $h1$ minibands.

We note that our calculated values of $B$ are, for structures having spatially indirect (type-II-like) carrier confinement, relatively high. At $T =$ 300 K and $n = p = 10^{17}$ cm$^{-3}$ our full SL calculation -- i.e.~including miniband dispersion and $q$-dependent optical matrix elements -- yields $B = 1.81 \times 10^{-11}$ cm$^{3}$ s$^{-1}$ for an InAs/GaSb SL having $t_{\scalebox{0.7}{\text{InAs}}} = t_{\scalebox{0.7}{\text{GaSb}}} = 2.14$ nm. This value is close to $B$ computed for bulk InAs from first principles, \cite{Hader_APL_2002} and compares favourably with that we have previously calculated for novel GaAs-based metamorphic InAs$_{1-x}$Sb$_{x}$/Al$_{x}$In$_{1-x}$As and InP-based pseudomorphic In$_{y}$Ga$_{1-y}$As$_{1-x}$Bi$_{x}$/In$_{0.53}$Ga$_{0.47}$As mid-infrared QWs having spatially direct (type-I) carrier confinement. \cite{Repiso_JPDAP_2019,Broderick_SST_2018} For example, this calculated value of $B$ is $\approx 31$\% of the value $B = 5.88 \times 10^{-11}$ cm$^{3}$ s$^{-1}$ we previously computed for a type-I InAs/Al$_{0.125}$In$_{0.875}$As metamorphic QW at the same temperature and carrier density. \cite{Repiso_JPDAP_2019} Despite having reduced electron-hole spatial overlap compared to structures possessing type-I band alignment, we conclude that partial delocalisation of near-zone-centre CB electrons in broken-gap InAs/GaSb SLs can produce a high radiative recombination rate suitable for applications in mid-infrared light-emitting devices. Our analysis therefore corroborates the experimental observation in Ref.~\onlinecite{Zhou_APL_2019} of high optical output power in prototype InAs/GaSb SL IC-LEDs.


\subsection{Unequal layer thickness: maximising the B coefficient at fixed emission wavelength}
\label{sec:results_unequal_thickness}

\begin{figure*}[t!]
	\includegraphics[width=1.00\textwidth]{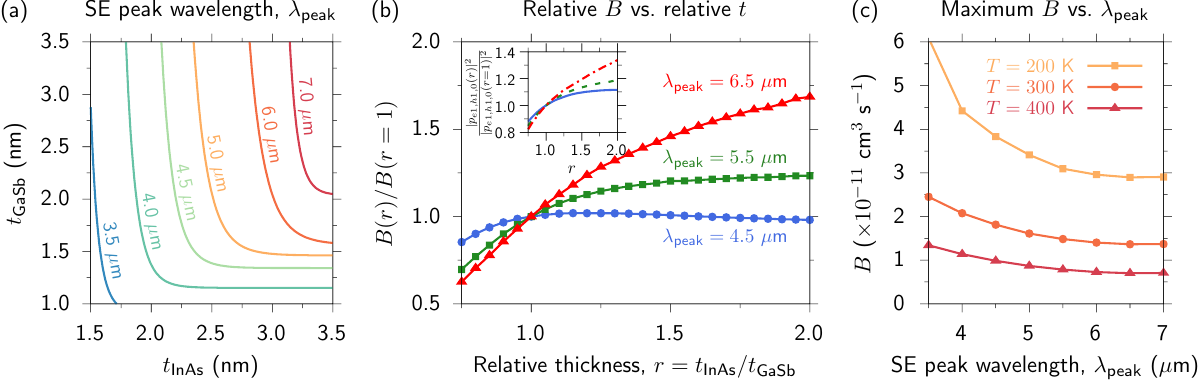}
	\caption{(a) Contour map denoting combinations of InAs and GaSb layer thicknesses $t_{\scalebox{0.7}{\text{InAs}}}$ and $t_{\scalebox{0.7}{\text{GaSb}}}$ that maintain fixed peak SE wavelengths $\lambda_{\scalebox{0.7}{\text{peak}}}$ ranging from 3.5 $\mu$m (blue) to 7 $\mu$m (red), at temperature $T =$ 300 K and carrier density $n = p = 10^{17}$ cm$^{-3}$. (b) Calculated variation of the radiative recombination coefficient $B$ as a function of the relative InAs and GaSb layer thickness $r = t_{\scalebox{0.7}{\text{InAs}}} / t_{\scalebox{0.7}{\text{GaSb}}}$, for $\lambda_{\scalebox{0.7}{\text{peak}}} =$ 4.5 $\mu$m (closed blue circles), 5.5 $\mu$m (closed green squares), and 6.5 $\mu$m (closed red triangles). For a given $\lambda_{\scalebox{0.7}{\text{peak}}}$, $B$ values are shown relative to their calculated value in an equal thickness structure having $r = 1$. Inset: relative squared TE-polarised optical matrix element vs.~$r$ for the same $\lambda_{\scalebox{0.7}{\text{peak}}}$ as in the main figure. (c) Maximised radiative recombination coefficient $B$ as a function of emission wavelength $\lambda_{\scalebox{0.7}{\text{peak}}}$, for temperature $T =$ 200 K (closed orange squares), 300 K (closed dark orange circles), and 400 K (closed red triangles).}
	\label{fig:fixed_wavelength}
\end{figure*}

Having elucidated the nature of the electronic structure and radiative recombination for an exemplar SL having equal InAs and GaSb layer thicknesses $t_{\scalebox{0.7}{\text{InAs}}}$ and $t_{\scalebox{0.7}{\text{GaSb}}}$, we now turn our attention to the maximisation of $B$ in InAs/GaSb SLs having unequal layer thicknesses, $t_{\scalebox{0.7}{\text{InAs}}} \neq t_{\scalebox{0.7}{\text{GaSb}}}$. For a chosen emission wavelength $\lambda_{\scalebox{0.7}{\text{peak}}}$ -- corresponding, as above, to the peak of the calculated SE spectrum at temperature $T = 300$ K for carrier density $n = p = 10^{17}$ cm$^{-3}$ -- we identify SLs that maintain fixed $\lambda_{\scalebox{0.7}{\text{peak}}}$ by systematically varying the relative thickness $r = t_{\scalebox{0.7}{\text{InAs}}}/t_{\scalebox{0.7}{\text{GaSb}}}$ of the InAs and GaSb layers. In the PWEM, a basis set consisting of $2M + 1$ plane waves possesses a minimum wavelength of $L/M$ (cf.~Sec.~\ref{sec:theory_electronic}). Since the SL period $L = t_{\scalebox{0.7}{\text{InAs}}} + t_{\scalebox{0.7}{\text{GaSb}}}$ can vary significantly with emission wavelength, and in fixed wavelength structures having $t_{\scalebox{0.7}{\text{InAs}}} \neq t_{\scalebox{0.7}{\text{GaSb}}}$, we treat SLs having different $L$ on an equal footing by varying $M$ so as to maintain $L/M$ as close as possible to the value $4.28/25$ nm $= 0.1712$ nm employed to analyse the equal layer thickness ($r = 1$) SL of Sec.~\ref{sec:results_equal_thickness} above. This ensures that the plane wave basis sets employed in the electronic structure calculations maintain equivalent real space resolution for the different structures considered. The results of this high-throughput structure search are summarised in Fig.~\ref{fig:fixed_wavelength}(a), which shows contours of constant $\lambda_{\scalebox{0.7}{\text{peak}}} ( t_{\scalebox{0.7}{\text{InAs}}}, t_{\scalebox{0.7}{\text{GaSb}}} )$ for emission wavelengths ranging from 3.5 $\mu$m (blue) to 7 $\mu$m (red).


Having identified these sets of fixed emission wavelength SLs, we then compute $B$ as a function of relative layer thickness $r = t_{\scalebox{0.7}{\text{InAs}}} / t_{\scalebox{0.7}{\text{GaSb}}}$ for each value of $\lambda_{\scalebox{0.7}{\text{peak}}}$. The results of these calculations are summarised in Fig.~\ref{fig:fixed_wavelength}(b) for $\lambda_{\scalebox{0.7}{\text{peak}}} = 4.5$ $\mu$m (closed blue circles), 5.5 $\mu$m (closed green squares), and 6.5 $\mu$m (closed red triangles) at $T = 300$ K. To highlight relative changes in $B$ and associated trends as a function of emission wavelength, each set of $B$ coefficients in Fig.~\ref{fig:fixed_wavelength}(b) is normalised to the $B$ coefficient calculated for an equal layer thickness ($r = 1$) SL at the corresponding value of $\lambda_{\scalebox{0.7}{\text{peak}}}$ -- i.e.~$B = 1$ for $r = 1$ at each fixed $\lambda_{\scalebox{0.7}{\text{peak}}}$. Figure~\ref{fig:fixed_wavelength}(b) therefore summarises the relative increase in $B$ that can in principle be achieved by varying the InAs and GaSb layer thicknesses in each period of an ideal InAs/GaSb SL. We note that quantitative values of layer thicknesses and $B$ coefficients are provided in Figs.~\ref{fig:fixed_wavelength}(a) and~\ref{fig:fixed_wavelength}(c), respectively.


Across the investigated wavelength range, our analysis predicts a strong reduction in $B$ for SLs having $r < 1$ (i.e.~$t_{\scalebox{0.7}{\text{InAs}}} < t_{\scalebox{0.7}{\text{GaSb}}}$). At $\lambda_{\scalebox{0.7}{\text{peak}}} = 4.5$ $\mu$m we find that $B$ is reduced by $\approx 15$\% when the relative thickness is reduced from 1 to 0.8, and that this reduction in $B$ is exacerbated with increasing emission wavelength (e.g.~decreasing by $\approx 40$\% when $r$ is reduced from 1 to 0.8 at $\lambda_{\scalebox{0.7}{\text{peak}}} = 6.5$ $\mu$m). The origin of this behaviour is straightforwardly understood: SLs in which the thickness of the hole-confining GaSb layers exceeds that of the electron-confining InAs layers reduce the ability of the $e1$ probability density to penetrate towards the centre of the GaSb CB barrier. This enhanced electron localisation acts to strongly reduce the $e1$-$h1$ optical matrix element (i.e.~the $e1$-$h1$ spatial overlap), and hence the radiative recombination rate. Our analysis therefore suggests that to maximise the radiative recombination rate, regardless of the choice of emission wavelength, priority should be given during growth to ensuring that the thickness of the electron-confining InAs layers is equal to or greater than that of the hole-confining GaSb layers. More generally, these results emphasise the importance of enhancing electron-hole spatial overlap in heterostructures having spatially indirect carrier confinement: the thickness of electron-confining layers, and the CB offset between those layers and the hole-confining layers, should be chosen to enhance electron delocalisation into the centre of the hole-confining layers in order to maximise $B$ (since such delocalisation is typically more readily achieved for low effective mass electrons than for high effective mass holes).

For SLs having $r > 1$ we find that the optimum relative layer thickness depends strongly on the chosen emission wavelength. Beginning with the $\lambda_{\scalebox{0.7}{\text{peak}}} =$ 4.5 $\mu$m results, we note that an equal layer thickness ($r = 1$) SL already presents a near-maximised $B$ coefficient. The maximum computed value of $B$ at $\lambda_{\scalebox{0.7}{\text{peak}}} = 4.5$ $\mu$m occurs for relative thickness $r = 1.15$, and is only 2.0\% larger than that calculated for the $r = 1$ SL. However, moving to longer wavelengths, our calculations identify significant potential to enhance $B$ by increasing the relative layer thickness $r$. Achieving longer emission wavelengths requires increased $t_{\scalebox{0.7}{\text{InAs}}}$, which in turn mandates a significant reduction in $t_{\scalebox{0.7}{\text{GaSb}}}$ to maintain fixed $\lambda_{\scalebox{0.7}{\text{peak}}}$ (cf.~Fig.~\ref{fig:fixed_wavelength}(a)). The resulting increase in $r$ allows for further penetration of the $e1$ probability density into GaSb layers, thereby increasing the $e1$-$h1$ spatial overlap and TE-polarised $e1$-$h1$ optical matrix element. This indicates that, for target emission wavelengths $> 5$ $\mu$m, the radiative recombination rate can be maximised by growing structures having relative layer thickness $r \approx$ 1.5 -- 2. Our analysis confirms that the evolution of the $e1$-$h1$ spatial overlap plays the dominant role in driving the calculated trends in $B$. This is summarised by the inset to Fig.~\ref{fig:fixed_wavelength}(b), which highlights that the relative increase of the squared $\mathbf{k} = 0$ TE-polarised $e1$-$h1$ optical matrix element squared, $\vert p_{\scalebox{0.7}{\text{$e1$-$h1$}}}^{(x,y)} \vert^{2}$, with $r$ -- for $\lambda_{\scalebox{0.7}{\text{peak}}} =$ 4.5 $\mu$m (solid blue line), 5.5 $\mu$m (dashed green line), and 6.5 $\mu$m (dash-dotted red line) -- closely tracks the calculated relative $B$ values. While this enhancement of $\vert p_{\scalebox{0.7}{\text{$e1$-$h1$}}}^{(x,y)} \vert^{2}$ accounts for the majority of the increase in $B$, we note that the reduction in $t_{\scalebox{0.7}{\text{GaSb}}}$ required to maintain fixed $\lambda_{\scalebox{0.7}{\text{peak}}}$ with increasing $r$ simultaneously acts to reduce the in-plane DOS close in energy to the VB maximum, which further contributes to enhancing the SE rate at fixed $T$ and carrier density. \cite{Ghiti_JAP_1992,Reilly_IEEEJQE_1994,Sweeney_JAP_2019}

Finally, Fig.~\ref{fig:fixed_wavelength}(c) shows the calculated maximum $B (r)$ as a function of $\lambda_{\scalebox{0.7}{\text{peak}}}$ for temperatures $T =$ 200 K (closed orange squares), 300 K (closed dark orange circles), and 400 K (closed red triangles). Here, having engineered the layer thicknesses to maximise the $e1$-$h1$ spatial overlap at each value of $\lambda_{\scalebox{0.7}{\text{peak}}}$, we note that the calculated values of $B$ track closely the $B \propto E_{g}$ scaling expected for an ideal bulk semiconductor (e.g.~reducing by close to a factor of 2 as the emission wavelength doubles from 3.5 to 7 $\mu$m). \cite{Lasher_PR_1964,Haug_APB_1987} Overall, our calculations suggest the potential to enhance the radiative recombination rate in InAs/GaSb SLs, through band structure engineering enabled by structural optimisation. This suggests that it may be possible to further improve on the optical output power and wallplug efficiency demonstrated in Ref.~\onlinecite{Zhou_APL_2019}, to produce high-performance light-emitters at technologically important mid-infrared wavelengths.


\section{Conclusions}
\label{sec:conclusions}

We have presented a theoretical analysis of radiative recombination in broken-gap InAs/GaSb SLs, based on multi-band $\mathbf{k} \cdot \mathbf{p}$ calculations of the SL electronic and optical properties. Exploiting the spatial periodicity of a plane wave basis set we established full $\mathbf{k}$-dependent calculations of SL band dispersion and eigenstates, with these quantities then used explicitly to compute the inter-band optical matrix elements, SE rate and radiative recombination coefficient $B$. Systematic optimisation of $B$ via high-throughput calculations provided guidelines for the growth of optimised SLs that maximise the radiative recombination rate across the application rich 3.5 -- 7 $\mu$m mid-infrared wavelength range.

The presence of spurious solutions in the calculated SL electronic properties was elucidated semi-analytically via the bulk complex band structure, and it was confirmed that the prescription of Ref.~\onlinecite{Foreman_PRB_1997} is sufficient to mitigate their impact in heterostructure calculations. The resultant renormalised InAs and GaSb band parameters were employed in full SL calculations, where it was demonstrated that the calculated temperature-dependent SE peak energy is in quantitative agreement with experimental EL measurements. Detailed analysis confirmed that miniband filling acts to decrease the temperature dependence of the SE peak energy compared to that of the fundamental zone-centre SL band gap. Extracted Varshni parameters for the temperature dependence of the SL SE peak energy were also found to be in quantitative agreement with additional experimental measurements.

Next, we considered InAs/GaSb SLs having equal InAs and GaSb layer thicknesses, and quantified the impact of miniband formation on the SE rate and $B$. Here, our calculations revealed the important roles played by (i) electron and hole localisation, (ii) carrier occupation of miniband states, and (iii) optical selection rules vs.~SL wave vector $q$ along a given pair of minibands. Firstly, it was shown that a combination of low InAs CB edge effective mass and GaSb layer thickness leads to partial electron delocalisation, allowing for high spatial overlap with hole states that are strongly localised in GaSb layers. This allows for high optical matrix elements, and hence for $B$ values that are large relative to other structures admitting spatially indirect carrier localisation. Secondly, the SL band structure was found to be characterised primarily by a mismatch between the dispersion of the lowest energy electron and highest energy hole minibands. The calculated electron minibands were found to be strongly dispersive such that, in the structure considered, only states out to $\vert q \vert \lesssim \frac{ \pi }{ 2L }$ have appreciable occupancy at room temperature. Conversely, the calculated hole minibands were found to be almost dispersionless, with holes then occupying miniband states throughout the entire SL BZ, $\vert q \vert \leq \frac{ \pi }{ L }$, at all temperatures. This mismatch in miniband dispersion prevents holes close to the SL BZ edge undergoing radiative recombination, so that miniband formation decreases both the magnitude of $B$ and its dependence on temperature. Finally, we undertook a systematic analysis of SLs having unequal InAs and GaSb layer thicknesses. We demonstrated that, at fixed emission wavelength: (i) at shorter wavelengths close to 3.5 $\mu$m $B$ is maximised in structures in which the InAs layer thickness is close to equal to the GaSb layer thickness, (ii) at longer wavelengths approaching 7 $\mu$m $B$ is maximised for InAs layer thickness $\approx 1.5$ -- $2 \; \times$ the GaSb layer thickness, and (iii) $B$ decreases rapidly if the InAs layer thickness is reduced below that of GaSb.

Strategies to increase the radiative recombination rate in SLs should incorporate reduction of the mismatch between the dispersion of the occupied electron and hole minibands, an approach akin to that deployed in the design of strained-layer QW lasers. Since the $q$-dependent optical matrix elements decrease strongly in magnitude with increasing $\vert q \vert$, our analysis suggests that it is beneficial to engineer the band structure such that the hole miniband dispersion is increased, so that fewer holes occupy states close to the SL BZ edge. Therefore, the ability to increase $B$ at fixed emission wavelength in SL structures beyond InAs/GaSb will rely in part upon the ability to match the electron and hole effective mass along the growth direction. This can potentially be achieved by exploiting layer thickness and/or strain engineering (enabled, e.g., via alloying and/or metamorphic growth).

Overall, our results emphasise the importance of considering the full $\mathbf{k}$-dependent miniband dispersion and eigenstates when analysing SL electronic and optical properties, and indicate that explicit calculations of this nature are essential to enable predictive analysis and quantitative interpretation of experimental data. While we have focused in this paper on InAs/GaSb as an archetypal mid-infrared SL system, we note that the established calculational framework is generally applicable to systems whose constituent material bulk band structures can be well described via a multi-band $\mathbf{k} \cdot \mathbf{p}$ Hamiltonian. This provides an accurate and computationally inexpensive approach to compute SL electronic and optical properties, providing a suitable platform for high-throughput calculations to support device design. Further analysis will be required to enable rigorous in silico optimisation of SL internal quantum efficiency, mandating accurate high-throughput calculation of the Auger recombination coefficient $C$ due to the importance of this non-radiative recombination process as a loss mechanism in narrow-gap heterostructures.


\section*{Acknowledgements}

This work was supported by the Irish Research Council (IRC; Government of Ireland Postgraduate Scholarship no.~GOIPG/2020/1252, held by C.M.), by the Science Foundation Ireland funded Irish Photonic Integration Centre (SFI-IPIC; grant no.~12/RC/2276\textunderscore{}P2), and by the European Commission's Horizon 2020 research and innovation programme via a Marie Sk\l{}odowska-Curie Actions Individual Fellowship (H2020-MSCA-IF; Global Fellowship ``SATORI'', grant agreement no.~101030927, held by C.A.B.). The authors thank Dr.~Qi Lu and Prof.~Anthony Krier (Lancaster University, U.K.) for providing access to the experimental data of Ref.~\onlinecite{Zhou_APL_2019}, and for useful discussions.


\section*{Data access statement}

The data associated with this work are openly accessible via Ref.~\onlinecite{sl_dataset}.



%

\end{document}


\preprint{AIP/123-QED}


\title[ ]{Supplemental Information: Theory and optimisation of radiative recombination in broken-gap InAs/GaSb superlattices}

\author{C\'{o}nal Murphy}
 \email{conalmurphy@umail.ucc.ie}
 \affiliation{Tyndall National Institute, University College Cork, Lee Maltings, Dyke Parade, Cork T12 R5CP, Ireland}
 \affiliation{School of Physics, University College Cork, Cork T12 YN60, Ireland}

\author{Eoin P.~O'Reilly}
 \affiliation{Tyndall National Institute, University College Cork, Lee Maltings, Dyke Parade, Cork T12 R5CP, Ireland}
 \affiliation{School of Physics, University College Cork, Cork T12 YN60, Ireland}

\author{Christopher A.~Broderick}
 \email{c.broderick@umail.ucc.ie}
 \affiliation{Tyndall National Institute, University College Cork, Lee Maltings, Dyke Parade, Cork T12 R5CP, Ireland}
 \affiliation{School of Physics, University College Cork, Cork T12 YN60, Ireland}

\date{\today}

\maketitle


\section{Complex band structure}
\label{sec:complex_band_structure}

We work with an 8-band \textbf{k}$\cdot$\textbf{p} Hamiltonian in the total angular momentum basis $\vert J; m_{J} \rangle$ of zone-centre Bloch states that diagonalise the spin-orbit interaction. \cite{Cohen_PRB_1990,Bahder_PRB_1990} With $k_{x} = k_{y} = 0$ the Hamiltonian block diagonalises such that the heavy-hole (HH) valence band (VB) is decoupled, with the dispersion of the light-hole (LH) and spin-split-off (SO) VBs and the lowest energy conduction band (CB) along [001] then described by the $3 \times 3$ Hamiltonian

\begin{widetext}
    \begin{equation}
	H_{3 \times 3} ( k_{z} ) = \left( \begin{array}{ccc}
            E_{g} + \frac{ \hbar^{2} }{ 2 m_{0} } \, s_{c} \, k_{z}^{2} & \sqrt{ \frac{2}{3} } P k_{z}                                                       & - \frac{1}{ \sqrt{3} } P k_{z} \\
             \sqrt{ \frac{2}{3} } P k_{z}                               & - \frac{ \hbar^{2} }{ 2 m_{0} } \left( \gamma_{\, 1} + 2 \gamma_{\, 2} \right) k_{z}^{2} & \sqrt{2} \frac{ \hbar^{2} }{ m_{0} } \gamma_{\, 2} \, k_{z}^{2} \\
            - \frac{1}{ \sqrt{3} } P k_{z}                              & \sqrt{2} \frac{ \hbar^{2} }{ m_{0} } \gamma_{\, 2} \, k_{z}^{2}                       & - \Delta_{0} - \frac{ \hbar^{2} }{ 2 m_{0} } \gamma_{\, 1} \, k_{z}^{2} \\
        \end{array} \right) \, \begin{array}{l} \vert \frac{1}{2} ; + \frac{1}{2} \rangle \\ ~ \vspace{-0.15cm} \\ \vert \frac{3}{2} ; + \frac{1}{2} \rangle \\ ~\vspace{-0.15cm} \\ \vert \frac{1}{2} ; + \frac{1}{2} \rangle \\ \end{array} \, ,
	   \label{eq:hamiltonian}
    \end{equation}
\end{widetext}

\noindent
where $E_{g}$ is the band gap, $\Delta_{0}$ is the VB spin-orbit splitting energy, and $P$ is the inter-band (Kane) momentum matrix element, related to the Kane parameter via $E_{P} = \frac{ 2m_{0}P^{2} }{ \hbar^{2} }$. The modified inverse CB edge effective mass $s_{c}$ is given by

\begin{equation}
    s_{c} = \frac{1}{ m_{c}^{\ast} } - \frac{ E_{P} }{3} \left( \frac{2}{ E_{g} } + \frac{1}{ E_{g} + \Delta_{0} } \right)
\end{equation}

\noindent
while the modified VB Luttinger parameters are defined as

\begin{eqnarray}
    \gamma_{\, 1} = \gamma_{\, 1}^{\text{\, L}} - \frac{ E_{P} }{ 3 E_{g} } \, , \\
    \gamma_{\, 2} = \gamma_{\, 2}^{\text{\, L}} - \frac{ E_{P} }{ 6 E_{g} } \, ,
\end{eqnarray}

\noindent
where $\gamma_{\, 1}^{\text{\, L}}$ and $\gamma_{\, 2}^{\text{\, L}}$ are the bare VB Luttinger parameters.

We compute the complex band structure admitted by Eq.~\eqref{eq:hamiltonian} using the eigenvalue method of Chang and Schulman.\cite{Chang_PRB_1982} We write Eq.~\eqref{eq:hamiltonian} as $H_{3 \times 3} ( k_{z} ) = H^{(0)} + H^{(1)} k_{z} + H^{(2)} k_{z}^{2}$, where $H^{(0)}$, $H^{(1)}$ and $H^{(2)}$ are the $3 \times 3$ coefficient matrices for terms in Eq.~\eqref{eq:hamiltonian} that are respectively independent of $k_{z}$, linear in $k_{z}$ and quadratic in $k_{z}$. Using these coefficient matrices we construct the $6 \times 6$ ``companion'' matrix to $H_{3 \times 3} ( k_{z} )$

\begin{equation}
     C ( E ) = \left( \begin{array}{cc}
     0                                    & I                          \\
     - ( H^{(2)} )^{-1} ( H^{(0)} - E I ) & - ( H^{(2)} )^{-1} H^{(1)} \\
     \end{array} \right) \, ,
     \label{eq:companion_matrix}
\end{equation}
\\
\noindent
where $I$ is the $3 \times 3$ identity matrix.

Diagonalising Eq.~\eqref{eq:companion_matrix} as a function of energy $E$ then yields the, in general, complex-valued wave vectors $k_{z,n} ( E, k_{x} = 0, k_{y} = 0 )$. The parameters used in our complex band structure calculations are as described for the SL calculations in the main text.


\section{Spurious solutions}
\label{sec:spurious_solutions}



Following the parameter renormalisation described in Sec.~II C of the main text, we note that it is not possible to conclude that the impact of spurious solutions on the SL electronic structure calculated using the 8-band \textbf{k}$\cdot$\textbf{p} Hamiltonian have been removed entirely. We will now demonstrate this explicitly. Firstly, we note that setting $s_{c} = 0$ allows for a discontinuity to occur at an InAs/GaSb interface in amplitude of the CB Bloch component of the envelope function. \cite{Foreman_PRB_1997} Omitting the SO band by setting the VB spin-orbit splitting $\Delta_{\scalebox{0.7}{\text{SO}}}$ equal to zero allows us to reduce $H_{3 \times 3}$ of Eq.~\eqref{eq:hamiltonian} to a 2-band Hamiltonian for the CB and LH states

\begin{equation}
H_{2 \times 2} ( k_{z} ) = \left( \begin{array}{cc}
        E_{c} + ak_{z}^{2} & Pk_{z} \\
        Pk_{z} & E_{v} - bk_{z}^{2} \\
    \end{array} \right) \, \begin{array}{l} \vert \frac{1}{2} ; + \frac{1}{2} \rangle \\ ~ \vspace{-0.15cm} \\ \vert \frac{3}{2} ; + \frac{1}{2} \rangle \\ \end{array} \, ,
	  \label{eq:hamiltonian_2x2}
\end{equation}

\noindent
where $a = \frac{\hbar^{2} s_{c}}{2 m_{0}}$ and $b = \frac{ \hbar^{2} }{ 2m_{0} } \left( \gamma_{\, 1} + 4\gamma_{\, 2} \right)$. We can use Eq.~\eqref{eq:hamiltonian_2x2} to understand the origin of the aforementioned discontinuity in the envelope function, which gives rise to the step-like behaviour of the $e1$ probability density observed in Fig.~2(a) of the main paper. We do so by diagonalising Eq.~\eqref{eq:hamiltonian_2x2} for small values of $a$ (equivalent to small values of $s_{c}$) and then analyse the evolution of its eigenstates as $a \to 0$. Evaluating the determinant $\vert H_{2 \times 2} - E \, I \vert = 0$ produces a quartic characteristic equation in $k_{z}$, for which the solutions for small $a$ are

\begin{widetext}
    \begin{equation}
        k_{z,\pm}^{2} (E) \approx \frac{ - \left( P^{2} + b \left( E_{c} - E \right) \right) \pm \sqrt{ \left( P^{2} + b \left( E_{c} - E \right) \right)^{2} -4ab \left( E_{c} - E \right) \left( E - E_{v} \right) } }{ 2ab } \, .
        \label{eq:k_solutions_small_a}
    \end{equation}
\end{widetext}

In addition to the usual Kane solution $k_{z,+}^{2}$, which describes the evanescent band linking the VB and CB edges, \cite{Das_PRA_2022} we also obtain a second solution $k_{z,-}^{2}$ whose value at small $a$ can be approximated as

\begin{equation}
    \kappa_{-}^{2} (E) \equiv - k_{z,-}^{2} (E) \approx \frac{ P^{2} + b \left( E_{c} - E \right) }{ ab } \, ,
    \label{eq:k_minus_small_a}
\end{equation}

\noindent
with $\kappa_{-}^{2} \to \infty$ as $a \to 0$. The corresponding eigenstates of $H_{2 \times 2}$ approach $| \psi_{+} \rangle = e^{\pm \kappa_{+}z} \vert \frac{1}{2}; + \frac{1}{2} \rangle$ as $a \to 0$, where $\vert \frac{1}{2}; + \frac{1}{2} \rangle$ is the $k_{z} = 0$ CB Bloch (basis) state. \cite{Bahder_PRB_1990} To calculate allowed energy states in a heterostructure grown along the $z$ direction, we replace $k_{z} \to - i \frac{ \text{d} }{ \text{d} z }$ in Eq.~\eqref{eq:hamiltonian_2x2}. \cite{Chen_PRB_1992} Writing this quantised Hamiltonian as

\begin{equation}
    \widehat{H} \psi = A \, \frac{ \text{d}^{2} \psi}{ \text{d} z^{2}} + B \, \frac{ \text{d} \psi}{dz} + C \psi \, ,
\end{equation}

\noindent
then the allowed solutions must satisfy the boundary conditions that both $\psi$ and $A \, \frac{ \text{d} \psi}{ \text{d} z } + \frac{1}{2} B \psi$ are continuous, \cite{Chen_PRB_1992} where

\begin{equation}
	A = \left( \begin{array}{cc}
            -a & 0 \\
            0 & b \\
        \end{array} \right) , \; B = \left( \begin{array}{cc}
            0 & iP \\
            -iP & 0 \\
        \end{array} \right) ,~\text{and} \; C = \left( \begin{array}{cc}
            E_{c} & 0 \\
            0 & E_{v} \\
        \end{array} \right) ,
	   \label{eq:A_matrix}
\end{equation}

\noindent
are the respective coefficient matrices of Eq.~\eqref{eq:hamiltonian_2x2} for terms that are quadratic in $k_{z}$, linear in $k_{z}$, and independent of $k_{z}$.

These boundary conditions appear to mandate that the CB component of $\psi$ must always be continuous. However, the term $A \, \frac{ \text{d} \psi}{ \text{d} z}$ can allow for a jump discontinuity in the amplitude of $\psi$ as $a \to 0$. For $E \approx E_{c}$ we have $\frac{ \text{d} \psi_{-} }{ \text{d} z } \approx \pm \frac{ P }{ \sqrt{ ab } } \psi_{-}$, so that the CB component of $A \, \frac{ \text{d} \psi}{ \text{d} z }$ can then vary as $- a \, \frac{ \text{d} \psi_{-}}{ \text{d} z } \approx \mp P \sqrt{ \frac{a}{b} } \, \psi_{-}$, which vanishes as $a \to 0$. This behaviour, discussed by Foreman in the $a = 0$ limit, \cite{Foreman_PRB_1997} accounts for the observed discontinuities in the $e1$ envelope function at InAs/GaSb interfaces in the full 8-band calculations (cf.~Fig.~2(a), main text).

We note that our calculated InAs and GaSb bulk complex band structures, using renormalised parameters with $s_{c} = 0$ (cf.~Fig.~1(b), main text), are in good quantitative agreement with the results of atomistic tight-binding calculations over the same ranges of energy and wave vector. \cite{Schulman_PRB_1985} States from outside of these ranges of energy and wave vector also contribute, albeit less strongly, to the determination of the variation of the electron envelope function with position in a full atomistic calculation. However, as described in the main text, our renormalised 8-band \textbf{k}$\cdot$\textbf{p} calculations capture the key features of the envelope functions and inter-band optical matrix elements observed in the full atomistic SL calculations of Refs.~\onlinecite{Kato_SM_2018} and~\onlinecite{Chang_PRB_1985}. This demonstrates that the impact of spurious solutions is sufficiently mitigated to validate our \textbf{k}$\cdot$\textbf{p} calculations of the InAs/GaSb SL electronic structure as a platform to predict and analyse trends in radiative recombination.


\section{Envelope function evolution along minibands}
\label{sec:envelope_function_evolution}

In the main text we noted that the optical (momentum) matrix element between the $\textbf{k}_{\parallel} = 0$ $e1$ and $h1$ SL eigenstates goes to zero as $q$ is varied from zero to $\frac{\pi}{L}$ (cf.~Fig.~3(a), main text). This behaviour was attributed to the introduction of a relative phase shift between the $e1$ and $h1$ envelope functions as the minibands are traversed in $q$. To illustrate this explicitly, we present here the calculated Bloch components of the $e1$ and $h1$ envelope functions at both $q = 0$ and $\frac{\pi}{L}$. We recall that since the HH band is decoupled at $\textbf{k}_{\parallel} = 0$, the $e1$ eigenstate consists of an admixture of CB, LH and SO Bloch character only, while the $h1$ eigenstate is purely HH-like.


\begin{figure*}[ht!]
	\includegraphics[width=0.98\textwidth]{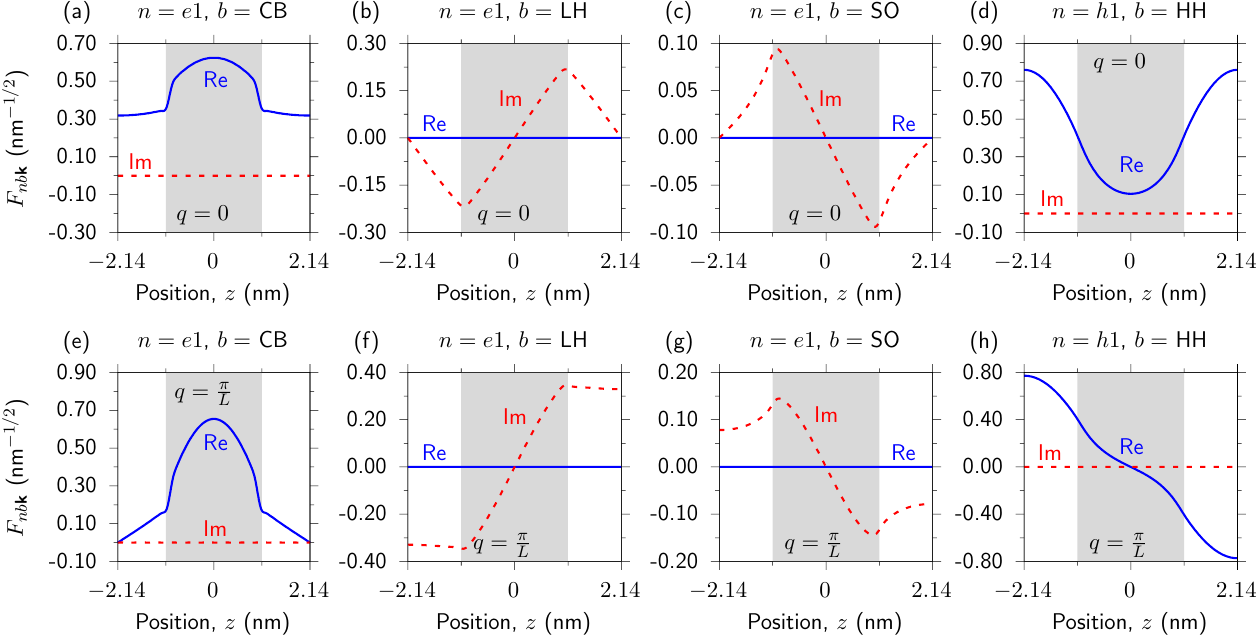}
	\caption{Top row: real (solid blue lines) and imaginary (dashed red lines) parts of the $q = 0$ envelope function Bloch components $F_{nb\textbf{k}} (z)$ for the $\textbf{k}_{\parallel} = 0$ $e1$ and $h1$ eigenstates of the exemplar equal layer thickness InAs/GaSb SL considered in the main text (having SL period $L = 2t = 4.28$ nm). (a) CB component of $e1$, (b) LH component of $e1$, (c) SO component of $e1$, and (d) HH component of $h1$. Bottom row: as in the top row, but for $q = \frac{\pi}{L}$. Shaded (unshaded) regions denote the electron-confining InAs (hole-confining GaSb) layers.}
	\label{fig:envelope_functions_e1}
\end{figure*}

The results of these calculations are summarised in Fig.~\ref{fig:envelope_functions_e1}, which shows the aforementioned non-zero envelope function Bloch components at $q = 0$ (top row) and $q = \frac{\pi}{L}$ (bottom row) for a single SL period. Panels (a), (b) and (c) in the top row, and (e), (f) and (g) in the bottom row, respectively show the CB, LH and SO components of the $e1$ eigenstate. Panels (d) and (h) show the HH component of the $h1$ eigenstate at $q = 0$ and $\frac{\pi}{L}$, respectively.

Considering first the Bloch components of the $e1$ eigenstate, at $q = 0$ the CB (LH and SO) component is purely real and even (imaginary and odd) about the centre of the InAs layer (shaded region) at $z = 0$, and remains so as $q$ changes from zero to $\frac{\pi}{L}$. For the $h1$ state, we observe changing phase and parity between $q = 0$ and $\frac{\pi}{L}$, with the HH Bloch component changing from being purely real and even about the centre of the InAs layer at $q = 0$ to purely real and odd about the centre of the InAs layer at $q = \frac{\pi}{L}$. This directly demonstrates that varying $q$ from zero to $\frac{\pi}{L}$ introduces a difference in parity between the $e1$ and $h1$ SL eigenstates at $\textbf{k}_{\parallel} = 0$. This results in the $e1$ and $h1$ eigenstates having equal (opposite) parity at $q = 0$ ($q = \frac{\pi}{L}$) which, combined with the respective $s$- and $p$-like symmetry of the CB and HH Bloch basis states $\vert \frac{1}{2}; \pm \frac{1}{2} \rangle$ and $\vert \frac{3}{2}; \pm \frac{3}{2} \rangle$, gives rise to the calculated reduction of the $\textbf{k}_{\parallel} = 0$ $e1$-$h1$ inter-band optical matrix element to zero at the SL Brillouin zone edge (cf.~Fig.~3(a), main text).

We note that the fact that this change in parity is restricted to the $h1$ eigenstate in Fig.~\ref{fig:envelope_functions_e1} is a consequence of the specific choice of calculational supercell. Here, we have utilised a supercell $z \in [ - \frac{L}{2}, \frac{L}{2} ]$ which is centred about the centre of the electron-confining InAs layer. Changing to, for example, a supercell with $z = 0$ set at the centre of the hole-confining GaSb layer instead produces a change in the phase and parity of the $e1$ eigenstate as $q$ varies from zero to $\frac{\pi}{L}$, with the parity of the $h1$ eigenstate then remaining fixed. This emphasises that it is the \textit{difference} in parity between the $e1$ and $h1$ eigenstates that determines their inter-band optical matrix element, irrespective of a given choice of calculational supercell.



%